\documentclass[sigconf]{acmart} 
\AtBeginDocument{%
  \providecommand\BibTeX{{%
    \normalfont B\kern-0.5em{\scshape i\kern-0.25em b}\kern-0.8em\TeX}}}

\usepackage{multirow}
\usepackage{makecell}

\setcopyright{acmcopyright}

\copyrightyear{2024}
\acmYear{2024}
\setcopyright{rightsretained}
\acmConference[CHI '24]{Proceedings of the CHI Conference on Human Factors in Computing Systems}{May 11--16, 2024}{Honolulu, HI, USA}
\acmBooktitle{Proceedings of the CHI Conference on Human Factors in Computing Systems (CHI '24), May 11--16, 2024, Honolulu, HI, USA}
\acmDOI{10.1145/3613904.3642660}
\acmISBN{979-8-4007-0330-0/24/05}

\begin{document}

\title[How Culture Shapes What People Want From AI]{How Culture Shapes What People Want From AI}



\author{Xiao Ge}
\authornote{These two authors contributed equally to this research.}
\affiliation{%
  \institution{Stanford University}
  \city{Stanford}
  \country{USA}}
\email{xiaog@stanford.edu}

\author{Chunchen Xu}
\authornotemark[1]
\affiliation{%
  \institution{Stanford University}
  \city{Stanford}
  \country{USA}
}
\email{cxu66@stanford.edu}

\author{Daigo Misaki}
\affiliation{%
  \institution{Kogakuin University}
  \city{Shinjuku, Tokyo}
  \country{Japan}}
\email{misaki@cc.kogakuin.ac.jp}

\author{Hazel Rose Markus}
\affiliation{%
  \institution{Stanford University}
  \city{Stanford}
  \country{USA}
}
\email{hmarkus@stanford.edu}

\author{Jeanne L Tsai}
\affiliation{%
  \institution{Stanford University}
  \city{Stanford}
  \country{USA}
}
\email{jltsai@stanford.edu}

\renewcommand{\shortauthors}{Ge, Xu, Misaki, Markus and Tsai}

\begin{abstract}
There is an urgent need to incorporate the perspectives of culturally diverse groups into AI developments. We present a novel conceptual framework for research that aims to expand, reimagine, and reground mainstream visions of AI using independent and interdependent cultural models of the self and the environment. Two survey studies support this framework and provide preliminary evidence that people apply their cultural models when imagining their ideal AI. Compared with European American respondents, Chinese respondents viewed it as less important to control AI and more important to connect with AI, and were more likely to prefer AI with capacities to influence. Reflecting both cultural models, findings from African American respondents resembled both European American and Chinese respondents. We discuss study limitations and future directions and highlight the need to develop culturally responsive and relevant AI to serve a broader segment of the world population.

\end{abstract}




\maketitle

\section{Introduction}

In the debris of a nuclear disaster, a sick and abandoned young woman awaits her death with her robot companion. The robot reads poems to her as the woman ponders death. This scene is from the 2015 Japanese film \textit{Sayonara} and depicts the near future of Japan. The faithful artificial intelligence (AI) companion was played by the humanoid robot, Geminoid F, created in Hiroshi Ishiguro’s research lab. With Geminoid F, Ishiguro wanted to show how ``android actors can express the vividness that cannot be expressed by humans'' \cite{Ishiguro2015}. The depiction of the robot in \textit{Sayonara} is strikingly different from most portrayals of AI companions in American films, which tend to be dystopian (e.g., \textit{The Terminator} and \textit{Her}). Indeed, a film critic in the U.S. described \textit{Sayonara} as ``[a] dreary study of human-robot relations [that] offers little to engage apart from its pretty scenery'' \cite{Debruge15}. Likewise, when \textit{Sayonara} was shown in theaters in the U.S.,  American audiences were ``astonished to see that the robots and androids were more sensitive than human beings, full of affection and at times capable of actions that were more thoughtful than their human counterparts'' \cite{Miyai2013}. 

We propose that these reactions in part reflect the fact that popular and literary depictions of AI in the U.S. tend to draw  clear, distinct, and hierarchical relationships between people and AI. The portrayal of AI as a cold, rational machine that operates in the background of—but also outside or external to—human society is consistent with cultural views that are prevalent in many European American cultural contexts. In these contexts, a person is seen as a free individual who makes choices and who controls or influences the objective world \cite{Markus2019,Morling2002}. The existence of AI entities that blur the distinction between emotional humans and dispassionate machines evokes terror and fear that these machines will control and dominate humans, as reflected in popular media in the U.S. and some European societies \cite{Bostrom2014,Friend2018}.

However, in many other sociocultural contexts, the foundational assumptions about individuals and their relationships with others and the environment are quite different. In East Asian contexts, for instance, individuals are seen as connected to and in reciprocal relationship with other people and the objective world. As Kohei Ogawa vividly explains, in Japan, “we can always see a deity inside an object” \cite{Zeeberg2020}. Within the multicultural U.S., many people endorse values of connectedness and construct agency differently than those common in European American contexts. For instance, Cree, Lakota, and Hawaiian Indigenous communities traditionally celebrate a sense of interconnection with all living beings and things. In these contexts, even engineered environments can be perceived as active, powerful, and full of life \cite{Kitano2007,Zeeberg2020}. Thus, based on these Indigenous epistemologies, Lewis and colleagues proposed to ``make kin'' with AI \cite{Lewis2018}. 
%

We propose that people's conceptions of AI—its purposes, forms, functions—and how AI should interact with humans are culturally variable. Yet, how cultures shape people's views of AI and its potential has not received significant empirical attention.
This lack of attention to culture in AI theory and design limits the space of the imaginary, and in particular, people's ideas about how humans and artificial agents might interact. Tapping into a variety of cultural ideas can catalyze new categories of AI, which may broaden the societal and environmental benefits of AI. Furthermore, in a multicultural world, many people identify with various mixes of cultures and/or interact with multiple, intersecting cultures in their daily lives, which should impact their views of AI. Thus, it is imperative for AI stakeholders to deepen their cultural understandings of human desires, beliefs and behaviors to make AI more relevant to a wider segment of the world population. 

As an initial step to uncover AI's cultural underpinnings, we drew upon theory and empirical research in cultural psychology regarding models of the self and the environment \cite{Markus1991,Fiske1998,Kitayama2005}. We conducted two exploratory survey studies on three cultural groups that have been observed in prior work to differ in their models of self: people in European American, African American, and Chinese contexts (hereafter referred to: \emph{European Americans, African Americans, and Chinese}). While people are influenced by their cultural contexts, these cultural groupings are not essentialized or monolithic categories. Thus, our objective is to examine general patterns that may emerge in spite of dynamic variations within each group and ongoing exchanges among cultural groups.

We first conducted a pilot study to confirm that Chinese, 
in comparison with European Americans and African Americans, were more likely to ideally view the environment as a source of influence on human behavior. Then, in the main study, we tested our hypotheses about how these cultural differences would be revealed in preferences for specific types of human-AI interaction. We found that, when imagining future ideal AI, European Americans preferred more control over AI and  preferred AI with fewer capacities to influence (i.e., wanted AI to have little autonomy, little spontaneity, and no emotion; to provide care but not need care from people, and to remain as abstract forms) than did Chinese. Chinese were more likely to seek connection with AI than were European Americans. African Americans' preferences reflected both cultural models, sharing features of both European American and Chinese ideals for human-AI interaction.

These strong preliminary findings are consistent with previous research demonstrating cultural differences in how people understand themselves and their relationships with their social and physical environments. Our findings suggest that people draw on these models when they imagine interacting with AI. 


Our work advances the human-computer interaction (HCI) field by:
\begin{itemize}

\item developing rigorous and systematic empirical approaches to examine people's culturally-shaped preferences regarding AI, 
\item illuminating the implicit and latent cultural assumptions about humans that are built into current models of human-computer interaction, and through this,
\item expanding current models of human-computer interaction to increase the potential of future technologies.
\end{itemize}



In the following sections, we will first review relevant work in HCI; introduce our theoretical framework; describe our studies and results; discuss implications of our research for AI development; and finally, discuss limitations of our work that generate directions for future research.

\section{Human-Computer Interaction Research on Culture}
\label{section:review}
Linxen and colleagues \cite{Linxen2021_weird} recently found that work in CHI is mostly based on Western, Educated, Industrialized, Rich, and Democratic (WEIRD) samples \cite{Henrich2010}. As reviewed in \citep{Kamppuri2006,Linxen2021_review}, between 1990 and 2006, only 28 out of 3286 papers (less than 1\%) published in CHI conference proceedings and four other prominent HCI publications \cite{Kamppuri2006} focused on culture. For the year 2010 and years 2016-2020, this percentage increased to a mere 1.9\%; only 1.6\% (N=58) of the CHI publications dealt explicitly with culture \cite{Linxen2021_review}. 

The cultural frameworks of Hofstede \cite{Hofstede2001,Hofstede1984} are the most often adopted ones in the reviewed studies. There are four general themes that characterize research on culture in HCI: (1) research that examines technological globalization, (2) research that aims to deepen situated understandings of human-computer interaction, (3) research that aims to expand design practices, and (4) last but not least, scholarly work that promotes equity. These themes are shown in Table \ref{tab:t.overview} . 

Most research falls under the first theme—studies that respond to technological globalization. As an example for cultural research on globally collaborative systems \cite{Scissors2011,Bauer2019,Chowdhury2011}, Scissors and colleagues studied how distributed teams from an international organization make use of real-time collaborative editing tools differently in Japan and the U.S. \cite{Scissors2011}. Many studies focus on how technology-mediated user behaviors vary across national cultures (e.g., use of assistive robots, social media tools) \cite{He2010,Lucas2018,Baughan2021,Korn2021,Li2010,Wang2010,Lee2014,Yang2013,Zhao2011,Cheng2021}. For instance, Cheng and colleagues \citep{Cheng2021} examined Facebook users across 18 countries and found experiences in social comparison on Facebook differ substantially by country.



The second theme describes research that aims to deepen situated understanding of human-machine interaction—e.g., situating the interaction in specific sociocultural contexts \cite{Marda2021,Taylor2011}, instead of focusing on context-free use \cite{Abrams2021,Chaves2021}. These efforts are primarily found in qualitative studies \cite{Alshehri2021,Marda2021,Devendorf2014}. For instance, through an ethnographic study, Marda and Narayan \cite{Marda2021} uncovered how implicit institutional cultures of Delhi policing contributed to the failures of New Delhi’s predictive policing system for crime control.


The third theme focuses on how cultural considerations have given rise to new design concepts and practices \cite{Heimgartner2013,Satchell2008,Rogers2012,Marcus2013}. For instance, culturally adaptive design concepts \cite{Kistler2012,Reinecke2011} were proposed to better respond to culture-specific human behaviors and accommodate cultural preferences. Some studies examined values of specific national contexts and how they are or could be embedded in the fabric of the local design practices \cite{Weng2019,Matsuzaki2016,Seymour2020,Hwang2018} or use cases \cite{Lu2021,Uriu2010,Yang2011,Gutierrez2016}. 

The fourth theme describes work that aims to promote equity. New policies and groups have been recently established to elevate diversity in scholarship \cite{Druin2018,Comber2020}. Theories have been proposed with enhanced cultural awareness \cite{Ogbonnaya2020} and/or through lenses of underrepresented populations \cite{McDonald2020,Soro2019}. In empirical work, the number of reported diversity dimensions (e.g., age, ethnicity and race, gender and sex, education, language) in CHI papers has increased \cite{Himmelsbach2019}, and more studies now intentionally include underrepresented populations \cite{Kumar2018}. For example, Woodruff and colleagues explored qualitatively how traditionally marginalized groups perceive algorithmic fairness \cite{Woodruff2018}; another recent study on resistance to AI decisions by Lee and Rich \cite{Lee2021} specifically focused on underrepresented group members in the U.S.

\begin{table}
  \caption{Overview of prior HCI research on culture: Four types of contributions}
  \label{tab:t.overview}
 \begin{tabular}{p{0.3\linewidth}  p{0.6\linewidth}}
    \toprule
    \ Theme & Example \\
    \midrule
    \\ Examining globalized technological developments, use and regulation
    &  Studies of the use of globally collaborative system \cite{Scissors2011,Bauer2019,Chowdhury2011}, and cross-nation difference in tech-mediated attitude and behavior  \cite{He2010,Lucas2018,Baughan2021,Korn2021,Li2010,Wang2010,Lee2014,Yang2013,Zhao2011,Cheng2021}  \\
     \\ Deepening situated understanding of HCI
     & Studies that situate interaction in specific sociocultural contexts \cite{Marda2021,Taylor2011,Abrams2021,Chaves2021}; Propositions to elevate nuanced cultural understandings \cite{Alshehri2021,Marda2021,Devendorf2014} \\
    \\Expanding design practice to consider culture-specific human behaviors and accommodate cultural preferences 
    &  Development of new design concepts and practices \cite{Heimgartner2013,Satchell2008,Rogers2012,Marcus2013,Kistler2012,Reinecke2011,Haddad2014,Kopecka2020}; Work to localize design practices \cite{Weng2019,Matsuzaki2016,Seymour2020,Hwang2018} or use cases \cite{Lu2021,Uriu2010,Yang2011,Gutierrez2016} \\
     \\ Promoting equity and increasing awareness of diversity, equity and inclusion in HCI
     & New policies, structural change in research \cite{Druin2018,Comber2020}; Studies that elevate representation of underrepresented populations through theory \cite{Ogbonnaya2020,McDonald2020,Soro2019_designing_past} and/or empirical work \cite{Himmelsbach2019,Kumar2018,Lee2021,Woodruff2018}\\
  \bottomrule
\end{tabular}
\end{table}

\subsection{Limitations}
Despite the importance of these efforts, scholars have noted some limitations of the existing research. First, many of the existing HCI studies of culture do not include the cross-cultural theories and research that have emerged from decades of research in the behavioral sciences \cite{Hekler2013,Shneiderman2002,Olson2003,Vatrapu2010,Linxen2021_review}. Culture is often viewed as a factor of usability, or a user interface design consideration. Cassell [15] observed that most discussion around cultural differences seems to scratch the surface and fall under “the rubric of user-modeling or personalization.”
 


Second, most studies tend to investigate the impact of technology on people and organizations in different countries (e.g., cultural differences in acceptability and usability of AI, or how AI impacts work). Almost no research \textit{flips the conversation} \cite{Sakura2021} to investigate how people's cultures shape their AI design practices and processes \cite{OLeary2019}, or how cultural ideas are reflected in and reinforced by AI products \cite{Cave2019,White2021,Sakura2021}. Only recently have researchers attempted to uncover AI's cultural underpinnings. For example, the Global AI Narratives is an on-going initiative that conducts workshops with local experts to examine how different societies perceive and envision AI \cite{GAIN}. 

The current research aims to build upon these initial efforts by examining how culture shapes people's preferences regarding AI, which may have important implications for the design of more culturally sensitive and relevant AI systems.



\section{Theoretical Framework and Focus of Current Research}
\subsection{Conception of Culture and Cultural Models}
We conceptualize \textit{culture} as ongoing patterns of visible and invisible meanings (e.g., ideas, representations, attitudes, mindsets and stereotypes), materialized interactions and practices (e.g., policies, social norms, languages), and products (e.g., media representations) that reflect and reinforce each other \cite{Adams2004}. These patterns change and evolve over time and therefore are not static entities \cite{Adams2001,Markus2019}.
In addition to geographically based characterizations, culture can include many other intersecting social distinctions, such as gender and sexuality, race and ethnicity, age, profession, social class, and more \cite{Markus2019}.


People think, feel, and act with reference to frameworks of tacit assumptions and meanings encoded in their local practices, social relationships, institutions, and artifacts. These assumptions and meanings are what researchers refer to as ``cultural models'' \cite{Fiske1998,Markus2019,Markus2016}. Cultural models can orient people's attention and feelings, lend meaning to their actions, and guide behavior. 

In this paper, we focus on these \emph{cultural models of the self and the environment}. We use this term to refer to patterns of ideas and practices about the ways people construe themselves, their particular social and physical environments, and the relations between the two. Consistent and robust empirical evidence has revealed two of a variety of normative ways in which people understand and define themselves in relation to their environment in various societies: independence and interdependence \cite{Markus2019}. We argue that these cultural models of self are germane to understanding different collective imaginations about AI.

\subsection{Cultural Models in European American, African American, and Chinese Contexts}
This paper examines three cultural groups: European Americans, African Americans, and Chinese based on previous evidence that they hold different cultural models of self \cite{Markus2019,Wang2004-fn,Kitayama2009-eu,Guan2015,Han2016,Brannon2015,Lyubansky2005,Walker2018}. Prior research has suggested that European Americans tend to view the self as independent, whereas Chinese tend to view the self as interdependent. African Americans tend to have both independent and interdependent views of the self \cite{Markus2019,Brannon2015,Lyubansky2005,Walker2018}.

In the \emph{independent} model of self, people primarily perceive themselves as unique individuals who are separate from others and from their physical and social environments \cite{Mazlish1993,Fiske1998}. People see their actions as mostly driven by internal goals, desires, and emotions \cite{Morling2002}, and they view the environment as an inert background for individuals to stand out and act on. People build a sense of self by expressing themselves and influencing or changing their environments to be consistent with their goals, desires, and emotions \cite{Dutta2011,tsai2007influence}.

In the \emph{interdependent} model of self, people tend to see themselves as fundamentally connected to others as well as to their physical and social environments. This has been traced to philosophical traditions that view people as closely connected to other people and as part of nature \cite{Markus1991,Savani2011}. In these contexts, people tend to respond to and act upon situational demands that are external to the person \cite{Kitayama2005}. As a result, their sense of the self depends critically on inferring the thoughts of others and adjusting their behavior to be consistent with others and their environments \cite{Kitayama2005}. 

These divergent cultural models are not inherent or essential but rather reflect the social and historical contexts that people inhabit \cite{Adams2001}.
Moreover, there are considerable differences among cultural sub-groups. For instance, within China, there are regions that are relatively more independent and relatively less interdependent than others \cite{Talhelm2014}. Overall, influenced by different cultural models, people locate the ``source of influence'' \cite{Markus2019} inside the self or outside in the environment to various degrees.  
To provide a more nuanced picture of within-nation variation, we go beyond European American-East Asian comparisons and include an African American group. Converging evidence suggests that African American models of self reflect a combination of independence and interdependence, and often vary depending on the prominence of mainstream European American and/or African American cultural ideas in a specific situation \cite{Brannon2015,Lyubansky2005,Walker2018}, explaining for seemingly inconsistent theorizing and findings in the past \cite{Oyserman2002-yd,Brannon2015}.
Because African American cultural contexts are characterized by multiple intersecting cultural models and can foster both independent and interdependent tendencies, African Americans may locate the source of influence in both the individual self and outside in the environment.

Prior work in cultural psychology has focused on how these models of self influence people’s relationships with other people. In HCI, some researchers have used cultural models of the self to examine how people interact with each other via various technologies, including Facebook use \cite{Kim2010}, distributed teamwork \cite{Scissors2011}, virtual agent-mediated conversation \cite{Lucas2018} and medical decision-making \cite{Lee2015}. However, few studies have examined how cultural models of self guide people’s  relations with elements of the non-human world (i.e., AI products and systems) \cite{Evers2008,Lee2012}. 
Due to limited empirical work comparing European American, African American, and Chinese models of self, we first ran a pilot study to test our theoretical framework.


\section{Hypothesis Development}

We propose that  cultural models of the self and the environment can predict what people want in their AI products and systems.

\subsection{Cultural Models Predict Ideal Human-AI Interactions}



Consider ambient intelligence, a field that originated in the West and that envisions a future in which physical environments continuously respond to people’s needs \cite{Remagnino2005,Aarts2006}. Core features of ambient intelligence include unobtrusive AI applications that operate quietly in the background while people in the foreground take control over their augmented environment  \cite{Remagnino2005,Aarts2009}. Thus, ambient intelligence is consistent with an independent model of the self, in which the environment is separate from and second to the individual \cite{Morling2002,Boiger2012}. By comparison, in cultural contexts characterized by an interdependent model of self, technological discourses do not necessarily assume a rigid boundary between humans and their technologies \cite{Jensen2013,White2021,Youichiro2020}. Instead, humans can coexist and blend with AI , which may be viewed as an extension of the natural world \cite{Youichiro2020}. In other words, in these contexts with independent models of self, artificial agents can play an active role in human social life rather than operate in the background \cite{MacDorman2009,Awad2018}. 

Thus, people can desire a more hierarchical, controlling relationship over AI, or a more equal and connecting relationship with AI. Here, we investigate whether people from different cultures vary in terms of their  desire for control over versus connection with AI.

Acknowledging the diversity and heterogeneity within each cultural group, we expect that, on average, Chinese would seek connection with AI more than European Americans would, due to a more interdependent orientation; European Americans would seek control over AI more than Chinese would, due to a more independent orientation; and African Americans would seek both control over and connection with AI because they are familiar with ideas and practices of both independence and interdependence. Although African Americans' bicultural orientation allows them to switch from one cultural identity to another, depending on the situation \cite{Brannon2015}, our studies did not focus on one cultural identity, so we expected that African American responses would fall between those of Chinese and European Americans.

This translated into the following hypotheses:

\begin{itemize}
\item H1a: European Americans will seek control over AI more than Chinese, whereas Chinese will seek connection with AI more than European Americans.

\item H1b: African Americans' desires to control and connect with AI will lie in between those of European Americans and Chinese.
\end{itemize}

Variation within each cultural group, while important, is beyond the scope of this paper.
\subsection{Cultural Models Shape Preferences Regarding AI's Capacities to Influence}
We were also interested in the specific characteristics that people prefer in AI. More specifically, to what extent do people want AI to have qualities that enable it to be an active form of influence, what we refer to as \textit{capacities to influence}? We propose that clues to this question can again be found by invoking cultural models of the self in relation to the environment.

In European American cultural contexts, 
people ideally see the self as a bigger source of influence than the environment. Consequently, in an ideal world, people should be more active, alive, capable, and in control than their environments, and people should aim to change their environments to be more consistent with their internal preferences, desires, and beliefs.  

In contrast, in Chinese contexts, the boundaries between individuals and their surrounding physical and social milieu are permeable and malleable. Thus, the environment and the self can share similar characteristics. Indeed, in Chinese contexts, people often conceptualize the environment as  encompassing them. As a result, they may prefer that the environment be more active, alive, capable and in control than the self. As a result,  individuals should aim to change their  internal preferences, desires, and beliefs to be consistent with their environments.

\begin{figure}[h]
 \centering
 \includegraphics[width=1\columnwidth]{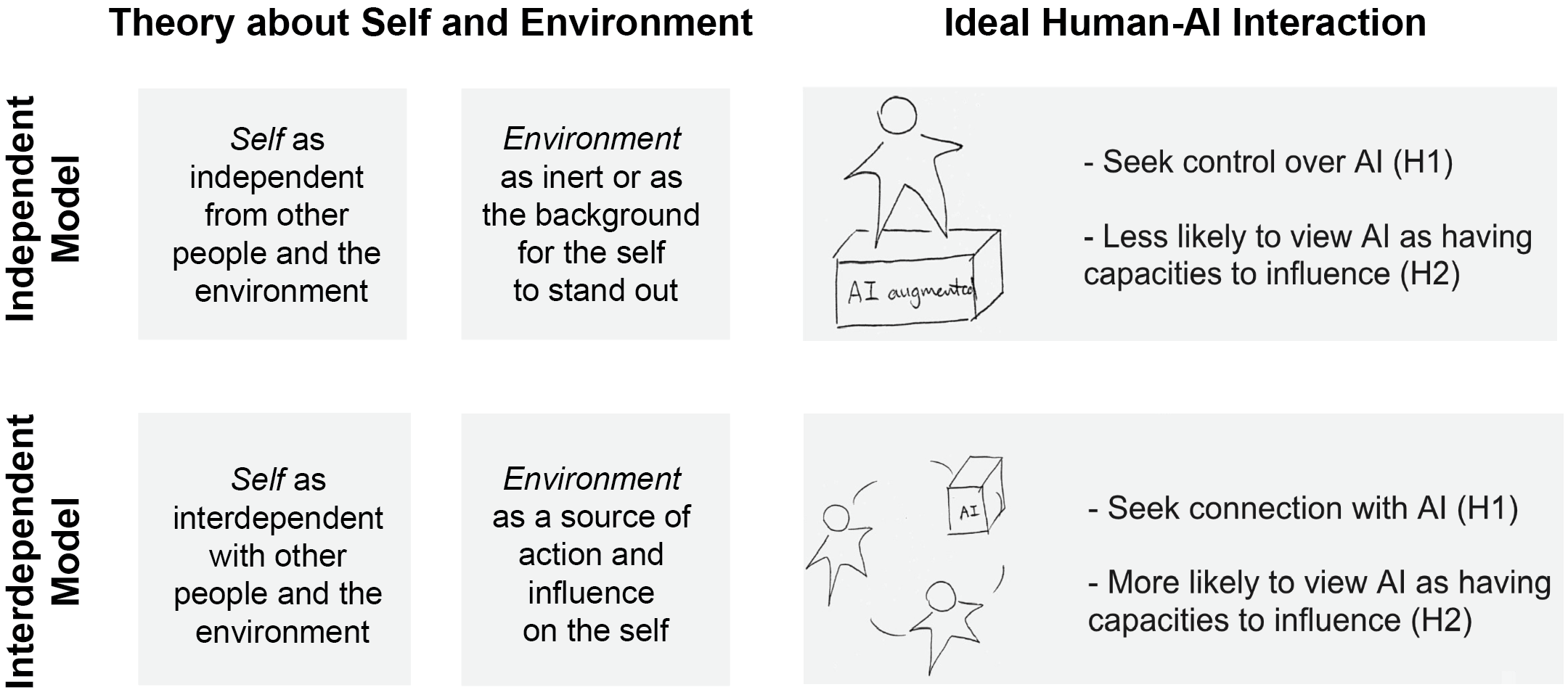}
 
\caption{Proposed links between cultural models of self and the environment and  ideal human-AI interaction
}
\label{fig:fig1 hypotheses}
\end{figure}

We argue that perceptions about the general environment can shed light on perceptions about AI products, which are part of the broader physical and social environment. If European Americans tend to place capacities to influence in individuals (versus the environment),  they should be less likely to think that AI could have qualities that would enable it to be an active force of influence. By comparison, if Chinese tend to place capacities to influence in environments (vs. the individual) more, they may be more willing to see AI as having capacities to influence. African Americans, with both cultural models, may hold views of AI's capacities to influence that fall in between those of European Americans and Chinese. We thus pose a second set of hypotheses:

\begin{itemize}
\item H2a: European Americans will be less likely to want AI to have capacities to influence than Chinese.

\item H2b: African Americans' desires for AI to have capacities to influence will fall between those of European Americans and Chinese.

\end{itemize}
Figure \ref{fig:fig1 hypotheses} summarizes our hypothesis development.


\subsection{Other Related Research}
While we mainly focus on cultural psychological research on independent and interdependent cultural models, our work is related to research on mind perception and anthropomorphism. 

We use \textit{capacities to influence} to describe an entity as a source of influence, action, and behavior  \cite{Markus1991}. Previous research suggests that people may perceive a given entity as having a \emph{mind} if they impart \emph{agency} and \emph{experience} to this entity \cite{Gray2007}. Similarly, the more an entity is perceived to have a ``mind,'' the more it has a capacity to influence. Thus, the items we developed to measure ``capacities to influence'' included ``feelings and emotions'' as well as ``initiative-taking,'' which  map onto the agency and experience dimensions of mind perception. However, we also depart from this body of work in one important way. Research on mind perception does not include aspects of agency that are relevant to both independent and interdependent cultural models. For example, the ``agency'' dimension of the mind perception focuses on self-control, which is an aspect of \emph{independent} agency \cite{Kitayama2005}. 
Therefore, we intentionally included aspects of \emph{interdependent} agency (e.g., harmony and flexibility) in our empirical studies. Furthermore, at a broader conceptual level, what is considered a ``mind'' can also be culturally variable. For instance, one recent study \cite{Spatola2022} found that participants from Germany and the U.S. shared a more anthropocentric view of the ``mind'' of a humanoid robot than did Korean and Japanese participants. Thus, we used capacities to influence (as opposed to perception of ``mind'') as a more general way to conceptualize and measure the extent to which people impart various characteristics to AI that may enable it to be an active force of influence.


Anthropomorphism \cite{Epley2007} is relevant to our research as well. Since we predicted that the people in interdependent cultural contexts (e.g., Chinese) are more likely to prefer AI as having capacities to influence, it follows that they are more likely to treat AI like  they treat humans. However, we recognize that some assumptions of anthropomorphism were primarily derived from the independent cultural models with a humanist tradition and an anthropocentric view on capacities to influence. This concept of "anthropomorphism" does not necessarily highlight the possibility that people in interdependent cultural contexts may treat elements of the environment (e.g., AI products) as active, alive, and intelligent without likening them to ``humans.'' In other words, people in these cultural contexts may have schemas other than ``humans'' to perceive elements of the environment that can influence them.

\section{Current Studies}



\begin{figure}[h]
 \centering
\includegraphics[width=.85\columnwidth]{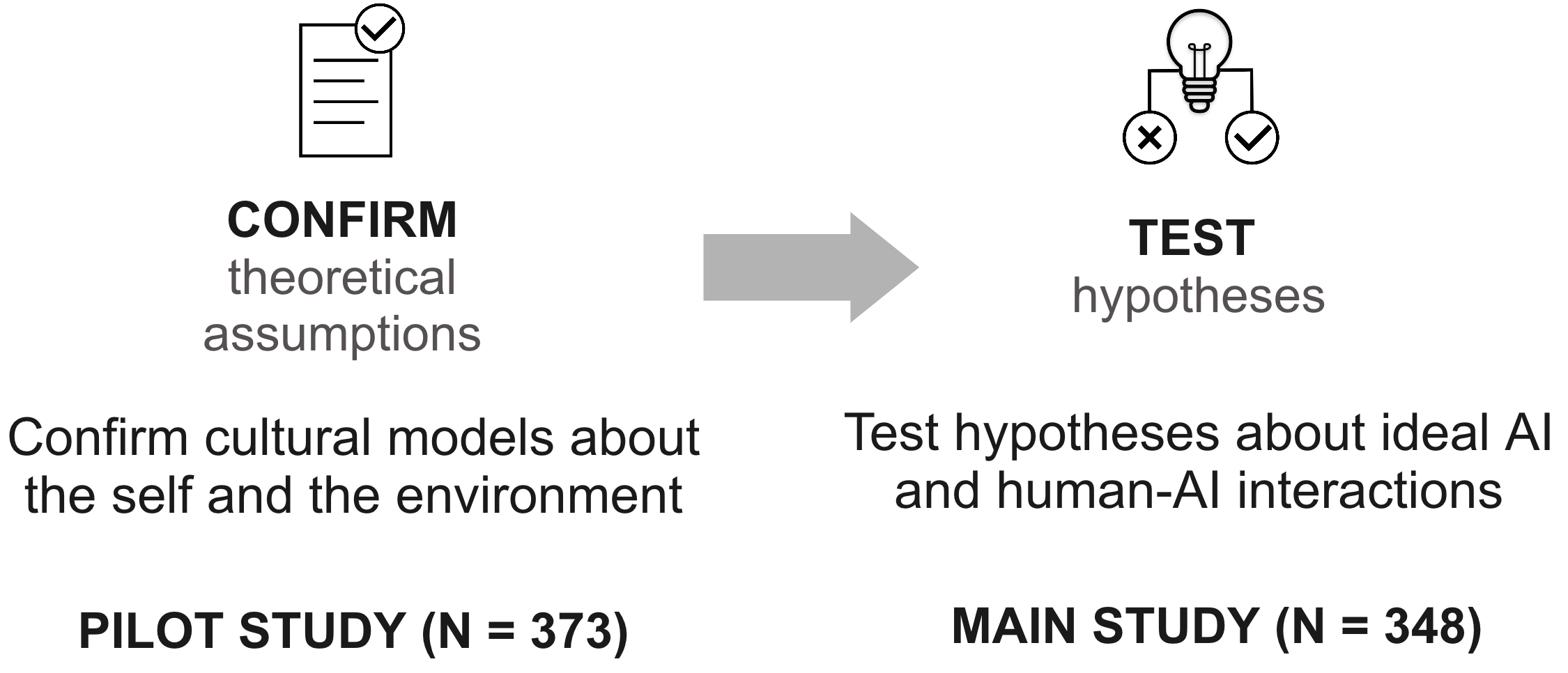}
\caption{Purposes of pilot and main studies}
\label{fig:fig2 steps}
\end{figure}

We designed two online survey studies. The Pilot Study aimed to confirm our theoretical assumptions that European Americans, African Americans, and Chinese have different cultural models of the self and the environment. After confirming these theoretical assumptions, we tested our hypotheses in the Main Study. Figure \ref{fig:fig2 steps} illustrates the two studies and their purposes. 
The surveys were administered in English for European and African American participants, and in Simplified Chinese for Chinese participants. Both surveys included attention and comprehension checks \citep{Abbey2017,Schwarz1999} to obtain better data quality. The following sections describe the methods and results of each study. 
Both studies were approved by the Stanford Institutional Review Board, and we only collected data from participants who consented to participate in the studies. All statistical analyses were conducted using R. 

To compare differences in means across the three groups, we performed one-way Analysis of Variance (ANOVA) or Multiple Analysis of Variance (MANOVA). We conducted post hoc Tukey's HSD tests to compare groups. In the Main Study, additional analyses were carried out to explore mediators of the observed effects (mediational analyses) and to control for other potential sources of variation (regression analyses). Data and surveys for the two studies are available at the Open Science
Framework (\url{https://osf.io/jn7a2/?view_only=e093b95b43b4474b928a632d8f11f1b6})

\section{Pilot Study}
 
\subsection{Methods}
\subsubsection{Participants}

We aimed to recruit  at least N = 100 participants from each group, based on typical sample sizes in previous cross-cultural research. Given the exploratory nature of this study, we erred on the side of recruiting more participants to account for those who might not pass attention checks. We recruited 179 European Americans and 198 African Americans via the survey platform CloudResearch. To select a relatively geographically diverse Chinese sample, we recruited 172 Chinese participants in Beijing who came from a variety of regions in China, via a Chinese survey platform. After removing participants who failed attention checks, we analyzed responses from a total of 131 European Americans (75 female, 56 male, M\textsubscript{age} = 43.05, SD\textsubscript{age} = 12.43), 121 African Americans (72 male, 48 female, 1 other, M\textsubscript{age} = 37.94, SD\textsubscript{age} =10.75) and 121 Chinese (64 female, 57 male, M\textsubscript{age} = 31.83, SD\textsubscript{age} =6.57).\footnote{Final analyses include responses from 121 (61.11\%) African Americans, 131 (73.18\%) European
Americans, and 121 (70.34\%) Chinese. 
The results were the same when we included all participants.


}


\subsubsection{Procedure}
Participants were introduced to the study and then answered questions (see \ref{Measures}) about their ideal models of the self and the environment.
Participants also reported their age, gender, and gross household income (before tax; measured on a 1-11 scale with each scale representing a range based on a reasonable income distribution in the U.S. and China, respectively).

\subsubsection{Measures}\label{Measures}
\begin{figure}[h]
 \centering
 \includegraphics[width=1\columnwidth]{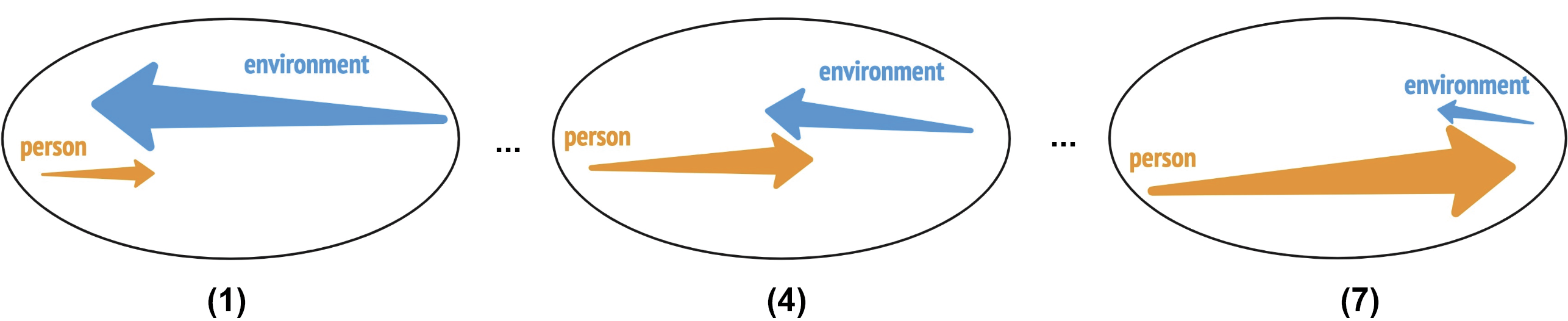}
\caption{The pictorial 7-pt. scale (only 1, 4 and 7 are presented for illustration purposes) used to measure the ideal level and direction of influence between the self and the environment in Pilot Study. 1 = ``The environment strongly influences the person,'' 7 = ``The person strongly influences the environment.''}
\label{fig:self-environment-influence-direction}
\end{figure}

While existing general measures about cultural differences (e.g., measures about individualism vs. collectivism \citep{Triandis1998}) are highly relevant, they do not directly tap into the specific cultural dimension we were interested in; namely, the relationship between the self and the environment. Therefore, we created seven different pictures representing the self and the environment  (\emph{The ideal level and direction of influence between the self and the environment}). Each picture varied in terms of the relative sizes of the two arrows to symbolize the extent to which the self influences the environment or vice versa, as shown in Figure \ref{fig:self-environment-influence-direction}. In each picture, we distinguished between the ``person'' and the ``environment'', which was broadly defined as ``factors that are external to a person in their daily life, including physical objects, other people, social norms, cultural ideas and so on.'' Participants were asked to choose the picture that best represented what they ideally wanted the relationship between people and their environments to be.




\subsection{Results}

\subsubsection{Cultural differences in the ideal level and direction of influence between the self and the environment}
\begin{figure}[h]
 \centering
 \includegraphics[width=.9\columnwidth]{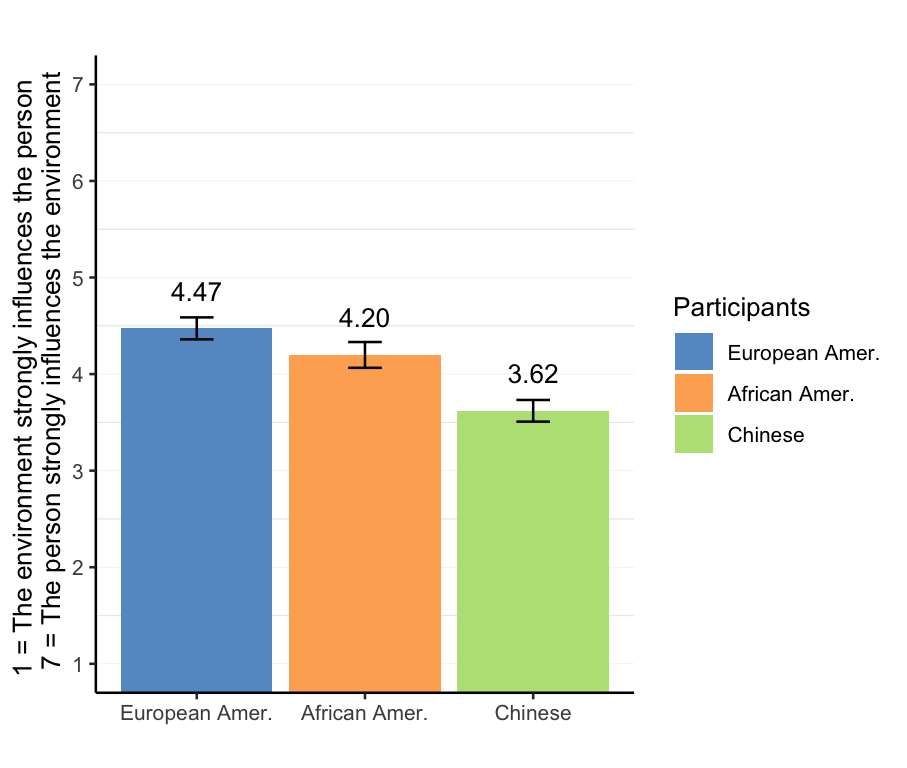}
\caption{Cultural differences in the ideal level and direction of influence between the self and the environment in Pilot Study. Error bars represent standard errors.}
\label{fig:result_pilot_study}
\end{figure}

A one-way ANOVA revealed a significant effect of cultural group, F(2,370) = 13.21, \emph{p} < .001, $\eta$\textsuperscript{2} = .07. Figure ~\ref{fig:result_pilot_study} illustrates the results. Tukey's HSD tests showed that compared with European Americans (M = 4.47, SD = 1.30) and African Americans (M = 4.20, SD = 1.47), Chinese (M = 3.62, SD = 1.23) were more likely to ideally want the environment to influence the person, \emph{p} < .001 and \emph{p} = .002, respectively. The difference between European and African Americans was in the predicted direction but was not significant, \emph{p} = .23. Table ~\ref{tbl.pilot.tukey} shows pairwise comparisons among the three groups.

\begin{table}[h]
\caption{\label{tbl.pilot.tukey}Tukey's HSD pairwise comparisons for analyses about cultural models across three groups in Pilot Study.}
{\begin{tabular}{lcccc}
\toprule
\multirow{2}{*}{Participants} & \multirow{2}{*}{Diff.} & \multicolumn{2}{c}{95\% CI} & \multirow{2}{*}{\textit{p} adj.} \\ \cmidrule{3-4}
                                 &       & \multicolumn{1}{c}{Lower} & \multicolumn{1}{c}{Upper} &         \\

\midrule
\multicolumn{1}{l}{\begin{tabular}[c]{@{}l@{}}Chinese vs. African Amer.   \end{tabular}}

& -0.58 & -0.98 & -0.17 & 0.002\\

\multicolumn{1}{l}{\begin{tabular}[c]{@{}l@{}}European vs. African Amer.   \end{tabular}}
& 0.27 & -0.12 & 0.67 & 0.23\\

\multicolumn{1}{l}{\begin{tabular}[c]{@{}l@{}}European Amer. vs. Chinese    \end{tabular}}
& 0.85 & 0.46 & 1.25 & $<$.001\\
\bottomrule
\addlinespace
\end{tabular}}

\end{table}








\subsubsection{Additional analyses}
In addition to their different cultural backgrounds, we found that the three samples also differed in terms of age (F(2,370) = 37.51, \emph{p} $<$ .001), gender ($\chi$\textsuperscript{2} (4, \emph{N} = 373) = 10.02, \emph{p} = .04), and annual household income (F(2,370) = 78.91, \emph{p} $<$ .001). Among these variables, age was correlated with ideal level and direction of influence.\footnote{We constructed one regression model predicting ideal relationship between the self and the environment with participants' cultural groupings while controlling for common demographic variables such as age, gender, and annual household income. We found that our effects remained statistically significant after common demographic variables were controlled for. However, it is worth noting that the model fit for this regression was not strong, and we suggest that future research expand on our findings and further explore how differences in demographic backgrounds might contribute to the observed effects.} Table ~\ref{tbl.pilot.corr} shows descriptive statistics and their correlations with ideal level and direction of influence. 

\begin{table}[h]
\caption{\label{tbl.pilot.corr}Descriptive statistics and correlations among variables in Pilot Study.}
 {\begin{tabular}{p{0.35\linewidth} p{0.1\linewidth}  p{0.1\linewidth} p{0.1\linewidth} p{0.05\linewidth}   p{0.05\linewidth}}
\toprule
  & 1 & 2 & 3 & $M$ & $SD$\\
\midrule
1. Ideal level and direction of influence btw. self and environment & - &  &  & 4.11 & 1.38\\
2. Gender\textsuperscript{a} & .01 & - &  & 0.50 & 0.50\\
3. Age & .17*** & -.12* & - & 37.75 & 11.25\\
4. Annu. household income & -.05 & .03 & -.24*** & 7.07 & 3.52\\
\bottomrule
\multicolumn{4}{l}{\begin{tabular}[l]{@{}l@{}}{\small Note: ``male'' was coded as 0 and ``female'' and ``other'' as 1.}  \end{tabular} }\\
&\multicolumn{5}{r}{\small  $^{*}$p$<$0.05; $^{**}$p$<$0.01; $^{***}$p$<$0.001} \\ 
\end{tabular}}

\end{table}

\subsection{{Pilot Study Summary and Discussion}}

In the Pilot Study, we found support for our theoretical assumptions that people from three different cultural groups have different  models of the self and environment. European Americans wanted people to influence their surrounding environments more. By comparison, Chinese wanted the environment to influence people more. African Americans' responses were similar to European Americans responses and were statistically different from those of Chinese. Based on these pilot findings, we proceeded to examine how these cultural models might affect people's views about ideal AI. 

\section{Main Study}

\subsection{Methods}
\subsubsection{Participants}

We recruited 142 European American, 123 African American and 175 Chinese participants via the same platforms, criteria, and procedures as in the Pilot Study. The sample size was estimated using G*Power. We expected a small to medium effect size for cultural differences in views of ideal AI based on the Pilot Study. With a significance criterion of $\alpha$ = .05 and 80\% statistical power, the minimum sample size needed for a small to medium effect size (Cohen's f = .20) is N = 246. Based on our previous studies, we estimated that if 65\% of participants passed attention checks, we would need to recruit N = 378 participants (N = 126 each group).\footnote{
We recruited more Chinese participants because we were unsure how many participants would pass attention checks.
} After removing people who failed attention checks, we analyzed responses from 127 European Americans (67 female, 59 male, and 1 other, M\textsubscript{age} = 40.43, SD\textsubscript{age} =11.50), 105 African Americans (50 female, 55 male, M\textsubscript{age} = 37, SD\textsubscript{age} =12.08) and 116 Chinese (35 female, 81 male, M\textsubscript{age} = 31.70, SD\textsubscript{age}= 6.91).\footnote{
Final analyses included responses from 105 (85.37\%) African Americans, 127 (89.43\%) European Americans and 116 (66.29\%) Chinese. The results were the same when we included all participants.
} 




\subsubsection{Procedure}
Participants first read a general description about AI: \emph{...AI can process information to recognize patterns, reproduce these patterns, make predictions, and help make decisions in different contexts. AI-driven technologies can use languages or other means to communicate with people. Some AI-driven technologies manifest an ability to learn, adjust and perform tasks in a changing environment.} Participants then read one of six scenarios of different AI applications, which were randomly assigned, based on the stimulus sampling method \cite{Monin2014-me}. The scenarios were not meant to be exhaustive of all AI applications but served as common ways in which AI is used (see Table~\ref{tab:t.study3.1}). After reading the scenario, participants answered questions about their ideal AI. Participants also reported their familiarity with AI, and answered the same demographic questions that were used in the Pilot Study (age, gender, and gross annual household income).

 \begin{table*}
   \caption{Scenarios of AI applications used  in Main Study.}
  \label{tab:t.study3.1}
   \begin{tabular}{p{0.2\linewidth} p{0.75\linewidth} }
    \toprule
    Scenario & Description\\
    \midrule
 Home management AI       & Imagine that in the future a home management AI is developed to communicate and manage people's needs at home and coordinate among many devices at their home. It makes customized predictions and decisions to improve home management. \\
 \makecell[tl]{Well-being \\ management AI}
   & Imagine that in the future a well-being management AI is developed to gather information about people's physical and mental health condition. It makes customized predictions and decisions to improve people’s well-being management.    \\
 Teamwork AI              & Imagine that in the future a teamwork AI is developed to collect information from team meetings and detect patterns of team interactions. It makes customized predictions and decisions to optimize team productivity and creativity.     \\
 Educational AI           & Imagine that in the future an educational AI is developed to analyze people's learning patterns and identify their strengths and weaknesses. It makes customized predictions and decisions to improve people's learning.                 \\
 Wildfire conservation AI & Imagine that in the future that a wildlife conservation AI is developed to monitor the population of endangered species. It makes predictions and decisions about how to conserve them.                                                 \\
 Manufacturing AI         & Imagine that in the future a manufacturing AI is developed to collect data during the process of manufacturing and detect errors on the assembly line. It makes predictions and decisions to improve manufacturing productivity.         \\
    \bottomrule
 \end{tabular}
\end{table*}

\subsubsection{Measures}

\textbf{Importance of having control over AI and connecting with AI}.
 We asked two questions to measure the importance of having control over AI and connecting with AI. The first question asked: ``How important is it for people to have control over what a/an [randomized contextual descriptor (e.g., educational, teamwork, wildlife conservation)] AI does?'' The second question asked, ``How important is it for people to have a sense of connection with a/an [randomized contextual descriptor (e.g., educational, teamwork, wildlife conservation)] AI?'' Participants used a 5-pt. scale (1 = ``Not important at all'' to  5 = ``Extremely important'') to answer both questions.

\textbf{Preference for AI's capacities to influence}. 
We developed a measure of participants' preference for AI's capacity to influence. After reading the scenarios, participants indicated their preferences for AI along nine different characteristics based on \emph{independent} and \emph{interdependent} cultural models of self (see Table 5). These nine characteristics incorporated concepts that are core to the field of HCI (e.g., AI's autonomy and its perceived emotionality), were related to AI's physicality and social presence based on ambient intelligence  \cite{Remagnino2005,Aarts2009}, and included characteristics that are central to interdependent models of self (emphasis on relationality, e.g., care for AI). Participants were asked to choose between two options (1 = ``low capacities to influence'' and 2 = ``high capacities to influence'') and then scores were summed across the six scenarios. The nine-item measure had good internal consistency ($\alpha$ = .86), and item-level descriptive statistics can be found in Table \ref{tbl.mainsty.descrp.ai.items}.

\begin{table*}

\begin{center}

\caption{\label{tbl.mainsty.descrp.ai.items} Descriptive statistics of the two-option nine-item measure about \textit{ideal} AI's capacities to influence among three groups in Main Study.}

\begin{tabular}{llcccccc}
\toprule
 \multirow{2}{*}{\begin{tabular}[c]{@{}l@{}} Low capacities to \\ influence (=1)   \end{tabular}} &
 \multirow{2}{*}{\begin{tabular}[c]{@{}l@{}} High capacities to \\ influence (=2)   \end{tabular}} & 
\multicolumn{2}{c}{\begin{tabular}[c]{@{}c@{}} European Amer. \end{tabular}} & \multicolumn{2}{c}{\begin{tabular}[c]{@{}c@{}}  African Amer.   \end{tabular}} & 
\multicolumn{2}{c}{\begin{tabular}[c]{@{}c@{}}  Chinese  \end{tabular}}  
 \\
\cline{3-8} 
 & & \multicolumn{1}{c}{\begin{tabular}[c]{@{}c@{}} M  \end{tabular}} &
 \multicolumn{1}{c}{\begin{tabular}[c]{@{}c@{}} SD  \end{tabular}} &
 \multicolumn{1}{c}{\begin{tabular}[c]{@{}c@{}} M  \end{tabular}} &
 \multicolumn{1}{c}{\begin{tabular}[c]{@{}c@{}} SD  \end{tabular}} &
 \multicolumn{1}{c}{\begin{tabular}[c]{@{}c@{}} M  \end{tabular}} &
 \multicolumn{1}{c}{\begin{tabular}[c]{@{}c@{}} SD  \end{tabular}} \\

\midrule
\multicolumn{1}{l}{\begin{tabular}[c]{@{}l@{}}AI provides care to but does not need care\\ from people.  \end{tabular}}
 & \multicolumn{1}{l}{\begin{tabular}[c]{@{}l@{}}AI provides care to but also needs care \\from people.   \end{tabular}}& 1.17 & 0.37 & 1.31 & 0.47 & 1.59 & 0.49 \\

\multicolumn{1}{l}{\begin{tabular}[c]{@{}l@{}}AI does not have feelings and emotions.  \end{tabular}}
 & \multicolumn{1}{l}{\begin{tabular}[c]{@{}l@{}}AI has feelings and emotions.   \end{tabular}}& 1.24 & 0.43 & 1.44 & 0.50 & 1.82 & 0.39 \\

 \multicolumn{1}{l}{\begin{tabular}[c]{@{}l@{}}AI remains an impersonal algorithm \\to perform tasks.  \end{tabular}}
 & \multicolumn{1}{l}{\begin{tabular}[c]{@{}l@{}}AI maintains a personal connection\\ with people.\end{tabular}}& 1.21 & 0.41 & 1.36 & 0.48 & 1.66 & 0.48 \\

 \multicolumn{1}{l}{\begin{tabular}[c]{@{}l@{}}AI operates unobtrusively in \\the background.  \end{tabular}}
 & \multicolumn{1}{l}{\begin{tabular}[c]{@{}l@{}}AI participates in social situations. \end{tabular}} & 1.20 & 0.41 & 1.35 & 0.48 & 1.66 & 0.48 \\

  \multicolumn{1}{l}{\begin{tabular}[c]{@{}l@{}}AI remains as an abstract algorithm \\whenever possible. \end{tabular}}
 & \multicolumn{1}{l}{\begin{tabular}[c]{@{}l@{}}AI has a tangible representation of \\its existence (e.g., a physical body)\\ whenever possible.  \end{tabular}} 
& 1.28 & 0.45 & 1.34 & 0.48 & 1.66 & 0.48 \\

 \multicolumn{1}{l}{\begin{tabular}[c]{@{}l@{}}AI has little autonomy. \end{tabular}}
 & \multicolumn{1}{l}{\begin{tabular}[c]{@{}l@{}}AI has autonomy.  \end{tabular}} 
 & 1.28 & 0.45 & 1.48 & 0.50 & 1.76 & 0.43 \\

\multicolumn{1}{l}{\begin{tabular}[c]{@{}l@{}}AI mainly preforms tasks that are  \\pre-planned by humans and has \\little spontaneity. \end{tabular}}
 & \multicolumn{1}{l}{\begin{tabular}[c]{@{}l@{}} AI has spontaneity when performing\\ tasks. \end{tabular}} 
 & 1.25 & 0.44 & 1.31 & 0.47 & 1.63 & 0.49 \\

 \multicolumn{1}{l}{\begin{tabular}[c]{@{}l@{}}AI behaves consistently across different\\ situations. \end{tabular}}
 & \multicolumn{1}{l}{\begin{tabular}[c]{@{}l@{}}AI behaves differently across different\\ situations. \end{tabular}}
& 1.33 & 0.47 & 1.40 & 0.49 & 1.89 & 0.32 \\

 \multicolumn{1}{l}{\begin{tabular}[c]{@{}l@{}}
AI interacts with people on terms \\ made by people. \end{tabular}}
 & \multicolumn{1}{l}{\begin{tabular}[c]{@{}l@{}}AI interacts with people on terms \\made by AI. \end{tabular}} & 1.20 & 0.41 & 1.20 & 0.40 & 1.52 & 0.50 \\

\bottomrule
\end{tabular}

\end{center}

\end{table*}

\textbf{Familiarity with AI use}. Participants were asked: ``How familiar are you with the use of Artificial Intelligence (AI)?'' on a 5-pt. scale (1 = ``None at all'' and 5 = ``A great deal'').

\subsection{Results}
Descriptive statistics and correlations among the variables are provided in Table~\ref{tbl.mainsty.cor}. In the following analyses, we collapsed across various AI scenarios.

\begin{table*}[tbp]

\begin{center}

\caption{\label{tbl.mainsty.cor}Descriptive statistics and correlation table in Main Study.}

\begin{tabular}{llllllllll}
\toprule
 & \multicolumn{1}{c}{1} & \multicolumn{1}{c}{2} & \multicolumn{1}{c}{3} & \multicolumn{1}{c}{4} & \multicolumn{1}{c}{5} & \multicolumn{1}{c}{6} & \multicolumn{1}{c}{$M$} & \multicolumn{1}{c}{$SD$} & \multicolumn{1}{c}{$Scale$}\\
\midrule
1. Imp. of having control over AI & - &  &  &  &  &  & 4.01 & 0.94 & 1-5\\
2. Imp. of connecting with AI & -.24*** & - &  &  &  &  & 3.17 & 1.21 & 1-5 \\
\makecell[tl]{3. Preferred capacities to influence in ideal AI} & -.48*** & .64*** & - &  &  &  & 1.42 & 0.34 & 1-2\\
4. Familiarity with AI use & .13* & .11* & .00 & - &  &  & 3.08 & 0.92 & 1-5\\
5. Gender & .00 & -.04 & -.07 & .11* & - &  & 0.44 & 0.50 &\\
6. Age & .11* & -.19*** & -.24*** & -.07 & .02 & - & 36.49 & 11.00 &\\
7. Annu. household income & -.09 & .18*** & .25*** & -.06 & -.15** & -.12* & 7.24 & 3.61& 1-11\\
\bottomrule 
 \multicolumn{10}{l}{\small Note: For the gender variable, ``male'' was coded as 0,  and ``female'' and ``other'' as 1.} \\ 
  \multicolumn{10}{r}{\small $^{*}$p$<$0.05; $^{**}$p$<$0.01; $^{***}$p$<$0.001} \\ 
\end{tabular}


\end{center}

\end{table*}

\subsubsection{Overall group differences.}\label{section:group_diff_all_dvs}

Due to dependencies among the three variables (importance of having control over AI, importance of connecting with AI, and AI's capacities to influence), we conducted a one-way MANOVA   \cite{warne2014primer}. We found a significant multivariate effect of culture, Pillai’s Trace = .34, F(6,345) = 23.73, \emph{p} <.001, partial $\eta$\textsubscript{p}\textsuperscript{2} = .17. Univariate analyses revealed that this was significant for all three variables: importance of having control over AI, F(2, 345) = 9.02, \emph{p} < .001, $\eta$\textsuperscript{2} = .05; importance of connecting with AI, F(2, 345) = 36.32, \emph{p} < .001, $\eta$\textsuperscript{2} = .17; ideal AI's capacities to influence, F(2, 345) = 83.39, \emph{p} < .001, $\eta$\textsuperscript{2} = .33. Table~\ref{tbl.sty2.contrlconn} shows a summary of the pairwise comparisons. 

\begin{table}[tbp]

\caption{\label{tbl.sty2.contrlconn}Tukey's HSD pairwise comparisons for the three DVs in Main Study.}

\begin{tabular}{llcccc}
\toprule
\multirow{2}{*}{Variable} & \multirow{2}{*}{\begin{tabular}[c]{@{}l@{}}Group \\comparison\end{tabular}} & \multirow{2}{*}{Diff.} & \multicolumn{2}{c}{95\% CI} & \multirow{2}{*}{\textit{p} adj.} \\ \cmidrule{4-5}
                                 &     &  & \multicolumn{1}{c}{Lower} & \multicolumn{1}{c}{Upper} &         \\
 
\midrule

 \multirow{3}{*}{\makecell[tl]{Having  \\control\\ over AI}}  & 
 \multicolumn{1}{l}{\begin{tabular}[c]{@{}l@{}} Chinese vs.\\ African Amer. \end{tabular}}
  & -0.35                    & -0.64                   & -0.06                   & 0.01                      \\
 &   
 \multicolumn{1}{l}{\begin{tabular}[c]{@{}l@{}}  
European vs.\\ African Amer.      \end{tabular}}    & 0.14                     & -0.15                   & 0.42                    & 0.50                      \\
 &  
 \multicolumn{1}{l}{\begin{tabular}[c]{@{}l@{}}  European Amer. \\vs. Chinese   \end{tabular}}
               & 0.49                     & 0.21                    & 0.77                    & <.001                      \\
\hline
\multirow{3}{*}{\makecell[tl]{Connecting \\with AI}}  & \multicolumn{1}{l}{\begin{tabular}[c]{@{}l@{}}  
Chinese vs.\\ African Amer.    \end{tabular}}     & 0.68                     & 0.34                    & 1.03                    & <.001                      \\
 &                                       
  \multicolumn{1}{l}{\begin{tabular}[c]{@{}l@{}}  
 European vs. \\African Amer.   \end{tabular}}              & -0.53                    & -0.87                   & -0.19                   & <.001                      \\
&                                
  \multicolumn{1}{l}{\begin{tabular}[c]{@{}l@{}}  
European Amer. \\vs. Chinese        \end{tabular}}                  & -1.20                    & -1.54                   & -0.87                   & <.001                      \\ 
\hline
 \multirow{3}{*}{\makecell[tl]{Preferred \\capacities  \\ to influence\\ in ideal AI}}  & 
 \multicolumn{1}{l}{\begin{tabular}[c]{@{}l@{}}  
 
 Chinese vs. \\African Amer.       \end{tabular}}     & 0.33                     & 0.24                    & 0.42                    & <.001                      \\
&                                            
 \multicolumn{1}{l}{\begin{tabular}[c]{@{}l@{}}  
European vs.\\African Amer.       \end{tabular}}   
       &             -0.12           & -0.20               &           -0.03           & .005                     \\
 &                       
  \multicolumn{1}{l}{\begin{tabular}[c]{@{}l@{}}  

European Amer. \\vs. Chinese          \end{tabular}}                     &              -0.45        &-0.53              &        -0.36               & <.001                      \\ 
\bottomrule
\end{tabular}

\end{table}

\subsubsection{Group differences in importance of having control over AI and connecting with AI}

\begin{figure}[]
  \centering
  \includegraphics[width=1\columnwidth]{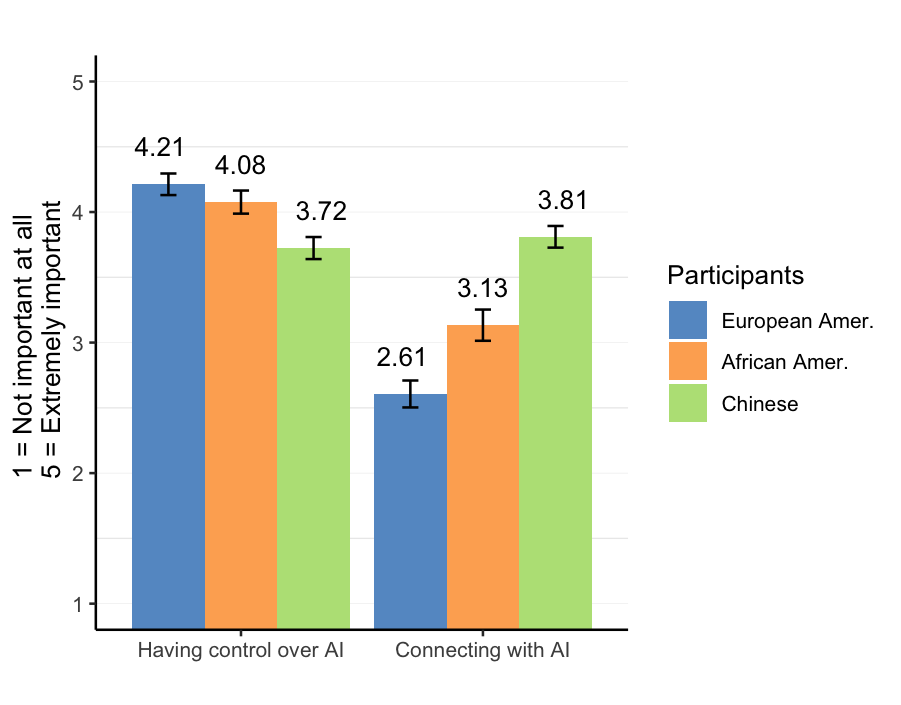}
  \caption{Importance of having control over AI and connecting with AI in Main Study, based on a 5-pt. scale. Error bars represent standard errors.}
  \Description{}
  \label{fig:f.contrlconn.sty3}
\end{figure}

Chinese rated controlling what AI does as less important (M = 3.72, SD = .91) than did both European Americans (M = 4.21, SD = .93) and African Americans (M = 4.08, SD = .91), adjusted \emph{p} < .001 and \emph{p} = .01, respectively. Chinese also rated having a sense of personal connection with AI as more important (M = 3.81, SD = .89) than did both European Americans (M = 2.61, SD = 1.16) and African Americans (M = 3.13, SD = 1.23), adjusted \emph{p}s < .001. Finally, African Americans rated it more important to connect with AI than did European Americans, adjusted \emph{p} <.001, but there was no significant difference between African Americans and European Americans in the importance of having control over AI, adjusted \emph{p} = .50. Figure~\ref{fig:f.contrlconn.sty3} illustrates the findings. 

\subsubsection{Group differences in preferences for ideal AI's capacities to influence}

\begin{figure}[]
  \centering
  \includegraphics[width=0.9\columnwidth]{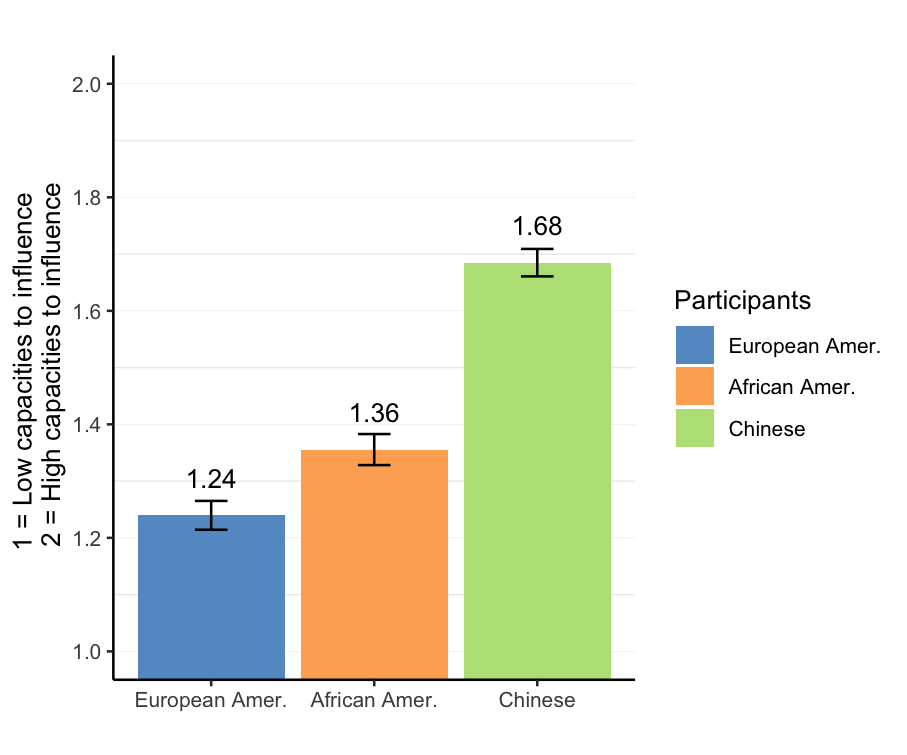}
  \caption{Preferred capacities to influence in ideal AI in Main Study, based on a 2-pt. scale. Error bars represent standard errors.}
  \Description{}
  \label{fig:f.study2}
\end{figure}

Compared with European Americans (M = 1.24, SD = .29), Chinese (M = 1.68, SD = .26) preferred AI to have higher capacities to influence, adjusted \emph{p} < .001. As predicted, African Americans' preference for ideal AI's capacities to influence (M = 1.36, SD = .28) fell between the above two groups\textemdash higher than that of European Americans', but lower than that of Chinese, adjusted \emph{p} = .005 and \emph{p} < .001. Figure~\ref{fig:f.study2} illustrates the finding.

\subsubsection{Exploratory mediation analyses}

Our theorizing implied that people imagine ideal AI's capacities to influence in a manner that is consistent with their preferred modes of human-AI interactions (i.e., having control over AI versus connecting with AI). Therefore, for exploratory purposes, we also examined if the importance placed on control versus connection mediated cultural differences in their preferences for AI's capacity to influence. 

To explore this possibility, we focused on European Americans and Chinese for comparison and constructed a 5000-iteration bootstrapped multiple-mediation model using the “Lavaan” package in R \cite{JSSv048i02}. We coded the European Americans as 0 and Chinese as 1, and collapsed across AI scenarios. We included the importance of having control over AI and connecting with AI as two mediators, and added the covariance of the two mediators to the model.

\begin{figure}[]
  \centering
  \includegraphics[width=1\columnwidth]{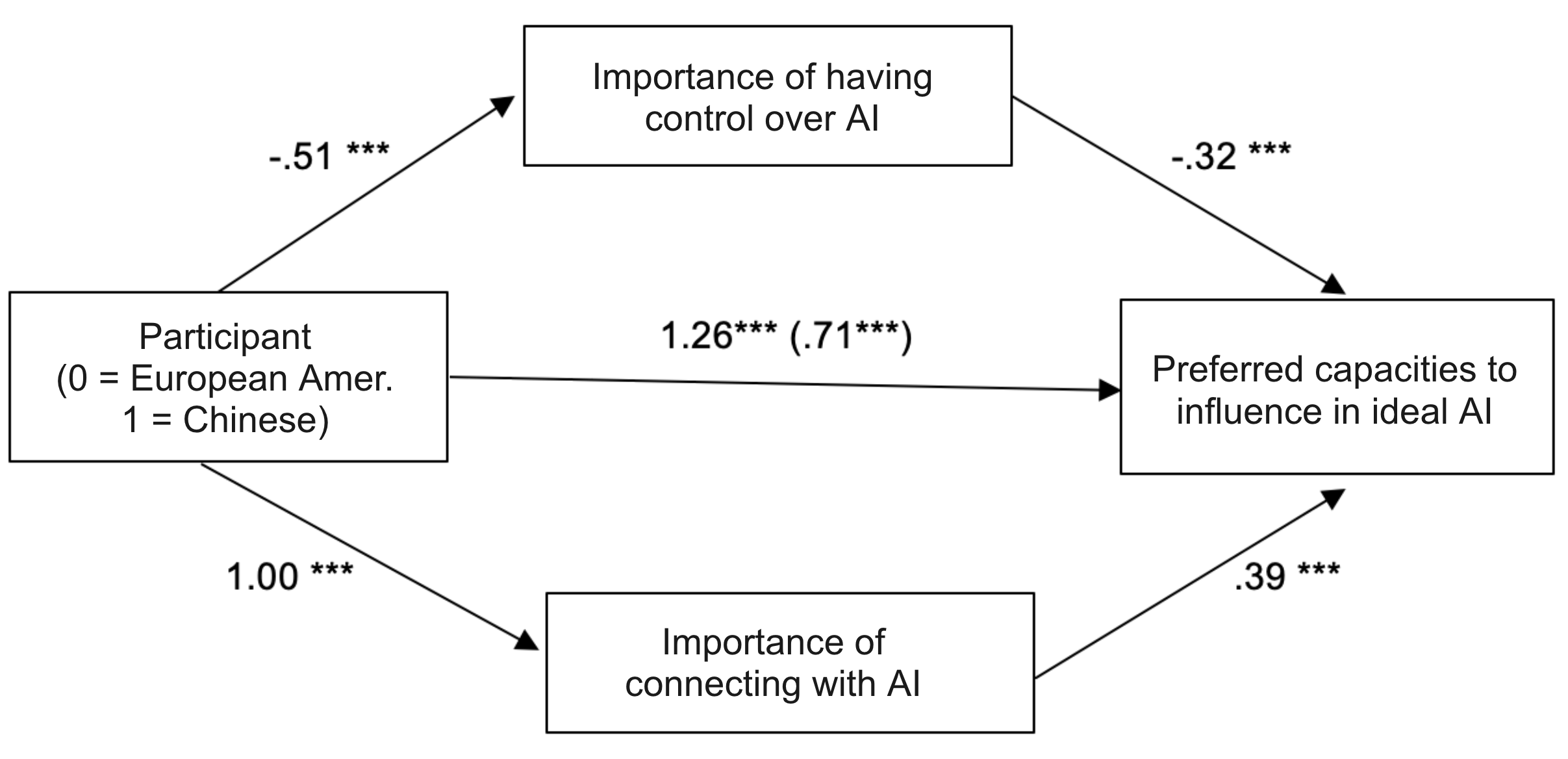}
  \caption{Cultural differences in preference for ideal Al's capacities to influence were mediated by cultural differences in the importance of having control over Al and connecting with Al in the Main Study. The direct effect of participants' cultural backgrounds on preferences for ideal AI’s capacities to influence is noted in parentheses.
  *\emph{p} < .05 **\emph{p} < .01, ***\emph{p} <. 001.}
  \Description{}
  \label{fig.mainsty.med}
\end{figure}

The bootstrapped standardized indirect effects for importance of control over AI was .17, \emph{CI\textsubscript{95}} = [.09, .25] and for importance of connection with AI was .39, \emph{CI\textsubscript{95}} = [.29, .52]. This suggested that these two mediators accounted for a significant amount of variance in the relationship between participant cultural backgrounds and their preferences for ideal AI's capacities to influence. After controlling for the indirect effects, the bootstrapped standardized direct effect remained significant at .71, \emph{CI\textsubscript{95}} = [.51, .92]. This indicated that the two mediators did not fully account for the observed effect. Overall, results from the mediation analyses provided some evidence that was consistent with our theorizing that cultural differences in preferences for ideal AI's capacities to influence were due to cultural differences in the importance of controlling vs. connecting with AI. Figure ~\ref{fig.mainsty.med} shows the mediation results.

\subsubsection{Additional analyses}

The cultural groups differed in terms of gender ($\chi$\textsuperscript{2} (4, \emph{N} = 348) = 15.62, \emph{p} = .004), age (F(2, 345) = 21.56, \emph{p} $<$ .001), and annual household income (F(2, 345) = 49.49, \emph{p} $<$ .001). There were more men in the Chinese samples (69.83\%), compared to the African American (52.38\%) and European American (46.46\%) samples. Chinese were also younger and reported higher positions on the income ladder than the other two groups. In addition, the three groups differed in their familiarity with AI, F(2, 345) = 9.34, \emph{p} < .001, $\eta$\textsuperscript{2} = .05. African Americans reported the highest levels of familiarity with AI use (M = 3.35, SD = .93), followed by European Americans (M = 3.09, SD = .88) and Chinese (M = 2.83, SD = .90). Tukey's HSD tests revealed that only the difference between African Americans and Chinese was significant, adjusted \emph{p} < .001, and the differences between the other two pairs were marginally significant, adjusted \emph{p}s = .07. 

To ensure that our findings held after controlling for these variables, we constructed three regression models predicting each of the DVs based on participants' cultural backgrounds (two dummy variables with Chinese as the reference group) while controlling for demographic variables and familiarity with AI use. Table \ref{tbl.mainsty.reg.contrl.conn.exist_updatedstars} illustrates the results of the additional analyses. Overall, we found that the observed cultural differences remained significant after controlling for gender, age, income, and familiarity with AI. It is worth noting that the first model has a fairly low adjusted R-squared value, suggesting that cultural background and other covariates are relatively less strong predictors than they are in the other two models. This result is consistent with our findings in section \ref{section:group_diff_all_dvs} showing that the effect size for cultural differences in the importance of control, $\eta$\textsuperscript{2} = .05, is smaller than effect sizes for the importance of connecting with AI and ideal AI's capacities to influence, $\eta$\textsuperscript{2}s =  .17 and .33, respectively.

Given that our dummy coding scheme did not allow an explicit comparison between African and European Americans, we constructed additional regression analyses to compare African and European Americans. The findings again remained robust even when we controlled for age, gender, income, and familiarity with AI.

\begin{table*}[!htbp] \centering 
  \caption{Regression for each of the DVs controlling for demographic variables and familiarity
with AI use in Main Study.} 
  \label{tbl.mainsty.reg.contrl.conn.exist_updatedstars} 
\begin{tabular}{lccc}
\toprule
 & \multicolumn{3}{c}{Dependent variable} \\ 
\cline{2-4} 
 & \multicolumn{1}{c}{\begin{tabular}[c]{@{}c@{}} Having control over AI   \end{tabular}}  & \multicolumn{1}{c}{\begin{tabular}[c]{@{}c@{}} Connecting with AI   \end{tabular}} & \multicolumn{1}{c}{\begin{tabular}[c]{@{}c@{}} Preferred capacities to influence in ideal AI  \end{tabular}} \\ 
 & (1) & (2) & (3)\\ 
\midrule
\multicolumn{1}{l}{\begin{tabular}[c]{@{}l@{}} African Amer. vs. Chinese \end{tabular}}   & 0.282$^{*}$ (0.001, 0.564) & $-$0.784$^{***}$ ($-$1.119, $-$0.450) & $-$0.343$^{***}$ ($-$0.427, $-$0.258) \\ 
\multicolumn{1}{l}{\begin{tabular}[c]{@{}l@{}}   European Amer. vs. Chinese \end{tabular}}  & 0.440$^{**}$ (0.167, 0.713) & $-$1.246$^{***}$ ($-$1.571, $-$0.922) & $-$0.447$^{***}$ ($-$0.529, $-$0.365) \\ 
Gender & $-$0.106 ($-$0.304, 0.093) & 0.082 ($-$0.154, 0.319) & 0.027 ($-$0.033, 0.087) \\ 
Age & 0.005 ($-$0.005, 0.014) & $-$0.004 ($-$0.015, 0.007) & $-$0.001 ($-$0.004, 0.001) \\ 
\multicolumn{1}{l}{\begin{tabular}[c]{@{}l@{}} Annu. household income  \end{tabular}}& $-$0.001 ($-$0.031, 0.030) & $-$0.0001 ($-$0.036, 0.036) & $-$0.0003 ($-$0.009, 0.009) \\ 
Familiarity with AI & 0.116$^{*}$ (0.008, 0.225) & 0.214$^{**}$ (0.085, 0.343) & 0.030$^{}$ ($-$0.003, 0.062) \\ 
Constant & 3.283$^{***}$ (2.728, 3.838) & 3.299$^{***}$ (2.638, 3.959) & 1.643$^{***}$ (1.476, 1.810) \\ 
\hline 
Observations & 348 & 348 & 348 \\ 
R$^{2}$ & 0.066 & 0.204 & 0.338 \\ 
Adjusted R$^{2}$ & 0.049 & 0.190 & 0.326 \\ 

\bottomrule
\multicolumn{4}{l}{\begin{tabular}[l]{@{}l@{}}{\small Note: Unstandardized coefficients are presented; 95\% confidence intervals (CI) are in parentheses.}  \end{tabular} }\\
& \multicolumn{3}{r}{\small  $^{*}$p$<$0.05; $^{**}$p$<$0.01; $^{***}$p$<$0.001} \\ 
\end{tabular} 
\end{table*}

\subsection{Study Summary and Discussion}
Findings from the Main Study overall support our hypotheses. Compared with European Americans, Chinese regarded it as less important to control AI but more important to have a sense of connection with AI (supporting H1a). European Americans preferred AI to have lesser capacities to influence (e.g., prefer AI to ideally \emph{not} have autonomy, spontaneity, emotion and \emph{not} need care from people) than did Chinese (supporting H2a). African Americans were similar to European Americans in the importance of having control over AI but were between European Americans and Chinese in terms of connecting with AI (partially supporting H1b). African Americans' preferences for ideal AI's capacities to influence were between European Americans and Chinese (supporting H2b). Notably, while Chinese placed the lowest importance on having control over AI, their average score (see Figure \ref{fig:f.contrlconn.sty3}) still reflects a desire to have control over AI.


\section{General Discussion}

Companies have been developing AI-based smart technologies at a remarkable rate, and yet, the foundational theories and practices that underlie these developments lack a rigorous consideration of cultural variation in what people want from AI. 
Here we propose that people draw on pre-existing cultural models about who they are and their relationship to the environment when thinking about their ideal relationship with AI. People from contexts that promote an interdependent model of self see the environment as an active source of influence on themselves, which leads them to seek connection with AI and to prefer AI with greater capacities to influence. By contrast, people from contexts that promote an independent model of self see themselves as controlling their environments, which leads them to seek control over AI and to prefer AI with low capacities to influence. 

In the Pilot Study, we demonstrated cultural differences in people's ideal relationship with the environment that reflected independent and interdependent models of the self, using a new measure.
Then, in the Main Study, we demonstrated cultural differences in European American, African American, and Chinese participants  desired interactions with AI that reflected these different ideal relationships with the environment, again using a new measure.  
 
These initial findings suggest that culture not only shapes people's conceptions of what it means to be human, but also what people desire in their interactions with AI. 



\subsection{Implications for Research on AI and Human-AI interactions}

\subsubsection{Rethinking control-based relationships in designing human-AI interaction}

Currently, many technological developments in the U.S. have focused on maintaining human autonomy and ensuring that humans ``have control over'' AI, a focus that we argue stems from an independent model of self. This view may fuel U.S. public distrust of AI companies and technologies more generally. Paradoxically, this view of AI might also limit AI developments and hide opportunities for developing different AI products that could benefit different societies across the globe.


Our findings suggest that the control-focused relationship is but one of the many possibilities for designing human-AI interaction. Findings from the Main Study suggests Chinese valued a connection-focused relationship with AI. They are not alone. Even in the U.S., there are examples of connection-focused human-AI exchanges such as the funeral held for a Martian rover at NASA \citep{Mack2019};  daily interactions with AI pet AIBO \citep{Friedman2003}; and improvisational performances by human and AI artists \citep{Savery2021}. Other scholars have advanced the idea of understanding and accommodating the agency of non-human things through a connection-focused lens \citep{Taylor2011}, including psychologist Kurt Lewin \citep{Lewin1999},  anthropologist Lucy Suchman \citep{suchman1987},  and historian and feminist Donna Haraway \citep{Haraway2006}. We suggest that connection-focused human-AI may fit the cultural values of a broader range of the world population and also address some of the societal problems associated with the current use of AI.

\subsubsection{Broadening the space of the imaginary for conceptualizing AI's characteristics}

Our theoretical framework and empirical findings challenge the default view of AI as impersonal systems that optimize, analyze, and repeat \citep{McCarthy2005}. 

Scholars within HCI \citep{Soro2019,Isbister2019} and beyond \citep{Lewis2018,Doctor2022} have called for collective reflection about the direction of future technologies. For instance, a few HCI scholars have similarly questioned the individualistic visions of technologies and proposed revolutions through ``suprahuman'' \citep{Isbister2019} and ``relational'' technology \citep{Soro2019}. Doctor and colleagues \cite{Doctor2022} suggest reconceptualizing intelligence based on Buddhist concepts, for instance.

We encourage more researchers to consider how we might design AI characteristics that reflect alternative visions. For instance, to design AI products that have high capacities to influence, we suggest that it would be worthwhile to experiment with designing AIs with ``needs'' to facilitate a more mutual and equal relationship between people and AI. Similarly, alternative AI systems could be foregrounded in social situations rather than operating unobtrusively in the background. This may require people to overcome their initial resistance and skepticism (Think of how  American audiences responded to \textit{Sayonara} in the opening example)—and at least consider other cultural views of AI.


\subsubsection{Re-evaluating how human-centeredness is conceived and implemented in AI developments}

Many AI technologies aim to put humans at the center of their design and fulfill human needs. However, ``human-centeredness'' is a deceptively simple idea. Humans within a society and across the globe differ in their values, beliefs, and behaviors, and therefore, human centered AI may look different depending on the cultural context.

We hope that the current studies will enable AI researchers, designers and policy-makers to reflect on the many implicit cultural assumptions about human behavior that have been built into the popular models of AI. We argue that it is important to consider culture at the initial design phase rather than only considering culture as an afterthought. 

Many technological organizations have departments of user research dedicated to human experience. However,  these organizations typically reward engineers and developers for developing cutting-edge technologies, rather than for designing for human needs that are culturally variable. It is our hope that the current work will also inspire reflections about these structural barriers at the organizational level. 

\subsection{Limitations and Future Directions}
Of course, our studies have a number of limitations. First, our findings were based on relatively small sample sizes, and there were important demographic differences among the three samples. Even though we analyzed potential confounding factors and checked the robustness of the results by controlling for demographic variables and familiarity with AI, there could be other latent, unobserved differences among the three samples that can provide alternative explanations for the observed effects. 
In future studies, we plan to include a more comprehensive set of control variables. For instance, it is plausible that individual differences in the need to belong \cite{baumeister1995need} could affect the extent to which people seek to connect with AI and control AI. Similarly, given there is ambiguity surrounding AI’s evolving roles in society, another meaningful variable could be tolerance of ambiguity \cite{furnham1995tolerance}. Finally, people's preferences may also be influenced by the degree to which they perceive that they may be replaced by AI in their jobs  \cite{brynjolfsson2017machine}.

Second, we developed new measures that were customized to capture people's ideal  models of self in relation to their environments and for understanding people's general preferences about AI and its capacities to influence. 
However, more work is needed to further establish these measures' reliability and validity. Future work interested in adopting the current measures should continue to refine them for their respective research contexts and study designs.


Third, our goal was to understand people's general views of and preferences for AI, so we defined the term AI in a general way and made it conceptually relatable through a variety of contextualized scenarios. Nonetheless, we did not study more specific AI products (e.g., chatbot or decision algorithms). 

Fourth, we relied on measures to capture people's reported preferences, but we do not know if these reported preferences predict people's actual interactions with AI. This is an important direction for future research. Design research methods that simulate realistic experiences, such as the wizard-of-oz method \cite{Steinfeld2009}, may be used to explore how people ideally want to interact with various AI products. 

Last but not least, more empirical evidence based on other culturally diverse samples is needed to test our theoretical claims. Future research can also examine other factors, such as gender, age, socioeconomic class, geography  \cite{Cohen2016,Gelfand2016,Markus2014}, as well as how they may change over time. This work should help reveal the heterogeneity within cultural groupings.

\section{Conclusions}




There is an urgent need to incorporate the preferences, imaginings, concerns, and creativity of diverse groups and communities into AI developments.
In this paper, we have begun to expand current visions of AI using different cultural models of the self and the environment. We provide theoretical frameworks and empirical tools for critically examining how culture shapes what people want in AI. The preliminary findings from the current  research contribute to our understanding of what  culturally responsive and relevant human-AI interactions might look like. We hope to inspire more basic research to understand the sociocultural construction of technology as well as more applied research to diversity AI design, with the ultimate goal of representing and serving a broader range of the world population.

\begin{acks}

Our research was supported by the Seed Grants offered by the Stanford Institute for Human-Centered Artificial Intelligence (HAI). We thank members of Cultural Collab (CCL) at Stanford Psychology Department for their valuable suggestions at various stages of this research project. We thank Cinoo Lee for providing thoughtful comments as we brainstormed survey questions and our participants for sharing their opinions that helped us better understand the cultural shaping of technological design and use.

\end{acks}

\bibliographystyle{ACM-Reference-Format}
\bibliography{chi23_cultureAI}


\begin{thebibliography}{126}


\ifx \showCODEN    \undefined \def \showCODEN     #1{\unskip}     \fi
\ifx \showDOI      \undefined \def \showDOI       #1{#1}\fi
\ifx \showISBNx    \undefined \def \showISBNx     #1{\unskip}     \fi
\ifx \showISBNxiii \undefined \def \showISBNxiii  #1{\unskip}     \fi
\ifx \showISSN     \undefined \def \showISSN      #1{\unskip}     \fi
\ifx \showLCCN     \undefined \def \showLCCN      #1{\unskip}     \fi
\ifx \shownote     \undefined \def \shownote      #1{#1}          \fi
\ifx \showarticletitle \undefined \def \showarticletitle #1{#1}   \fi
\ifx \showURL      \undefined \def \showURL       {\relax}        \fi
\providecommand\bibfield[2]{#2}
\providecommand\bibinfo[2]{#2}
\providecommand\natexlab[1]{#1}
\providecommand\showeprint[2][]{arXiv:#2}

\bibitem[Aarts and de~Ruyter(2009)]%
        {Aarts2009}
\bibfield{author}{\bibinfo{person}{Emile Aarts} {and} \bibinfo{person}{Boris de Ruyter}.} \bibinfo{year}{2009}\natexlab{}.
\newblock \showarticletitle{New research perspectives on Ambient Intelligence}.
\newblock \bibinfo{journal}{\emph{Journal of Ambient Intelligence and Smart Environments}}  \bibinfo{volume}{1} (\bibinfo{year}{2009}), \bibinfo{pages}{5--14}.
\newblock
Issue 1.
\showISSN{18761364}
\urldef\tempurl%
\url{https://doi.org/10.3233/AIS-2009-0001}
\showDOI{\tempurl}


\bibitem[Abbey and Meloy(2017)]%
        {Abbey2017}
\bibfield{author}{\bibinfo{person}{James~D Abbey} {and} \bibinfo{person}{Margaret~G Meloy}.} \bibinfo{year}{2017}\natexlab{}.
\newblock \showarticletitle{Attention by design: Using attention checks to detect inattentive respondents and improve data quality}.
\newblock \bibinfo{journal}{\emph{J. Oper. Manage.}} \bibinfo{volume}{53-56}, \bibinfo{number}{1} (\bibinfo{date}{Nov.} \bibinfo{year}{2017}), \bibinfo{pages}{63--70}.
\newblock


\bibitem[Abrams et~al\mbox{.}(2021)]%
        {Abrams2021}
\bibfield{author}{\bibinfo{person}{Anna M.~H. Abrams}, \bibinfo{person}{Pia S.~C. Dautzenberg}, \bibinfo{person}{Carla Jakobowsky}, \bibinfo{person}{Stefan Ladwig}, {and} \bibinfo{person}{Astrid~M. Rosenthal-von~der P\"{u}tten}.} \bibinfo{year}{2021}\natexlab{}.
\newblock \showarticletitle{A Theoretical and Empirical Reflection on Technology Acceptance Models for Autonomous Delivery Robots}. In \bibinfo{booktitle}{\emph{Proceedings of the 2021 ACM/IEEE International Conference on Human-Robot Interaction}} (Boulder, CO, USA) \emph{(\bibinfo{series}{HRI '21})}. \bibinfo{publisher}{Association for Computing Machinery}, \bibinfo{address}{New York, NY, USA}, \bibinfo{pages}{272–280}.
\newblock
\showISBNx{9781450382892}
\urldef\tempurl%
\url{https://doi.org/10.1145/3434073.3444662}
\showDOI{\tempurl}


\bibitem[Adams and Markus(2001)]%
        {Adams2001}
\bibfield{author}{\bibinfo{person}{Glenn Adams} {and} \bibinfo{person}{Hazel~Rose Markus}.} \bibinfo{year}{2001}\natexlab{}.
\newblock \showarticletitle{Culture As Patterns: An Alternative Approach to the Problem of Reification}.
\newblock \bibinfo{journal}{\emph{Culture \& Psychology}} \bibinfo{volume}{7}, \bibinfo{number}{3} (\bibinfo{year}{2001}), \bibinfo{pages}{283--296}.
\newblock
\urldef\tempurl%
\url{https://doi.org/10.1177/1354067X0173002}
\showDOI{\tempurl}
\showeprint{https://doi.org/10.1177/1354067X0173002}


\bibitem[Adams and Markus(2004)]%
        {Adams2004}
\bibfield{author}{\bibinfo{person}{Glenn Adams} {and} \bibinfo{person}{Hazel~Rose Markus}.} \bibinfo{year}{2004}\natexlab{}.
\newblock \bibinfo{booktitle}{\emph{Toward a Conception of Culture Suitable for a Social Psychology of Culture.}}
\newblock \bibinfo{publisher}{Lawrence Erlbaum Associates Publishers}, \bibinfo{pages}{335--360}.
\newblock
\showISBNx{0-8058-3839-2 (Hardcover); 0-8058-3840-6 (Paperback)}


\bibitem[Alshehri et~al\mbox{.}(2021)]%
        {Alshehri2021}
\bibfield{author}{\bibinfo{person}{Taghreed Alshehri}, \bibinfo{person}{Norah Abokhodair}, \bibinfo{person}{Reuben Kirkham}, {and} \bibinfo{person}{Patrick Olivier}.} \bibinfo{year}{2021}\natexlab{}.
\newblock \showarticletitle{Qualitative Secondary Analysis as an Alternative Approach for Cross-Cultural Design: A Case Study with Saudi Transnationals}. In \bibinfo{booktitle}{\emph{Proceedings of the 2021 CHI Conference on Human Factors in Computing Systems}} (Yokohama, Japan) \emph{(\bibinfo{series}{CHI '21})}. \bibinfo{publisher}{Association for Computing Machinery}, \bibinfo{address}{New York, NY, USA}, Article \bibinfo{articleno}{144}, \bibinfo{numpages}{15}~pages.
\newblock
\showISBNx{9781450380966}
\urldef\tempurl%
\url{https://doi.org/10.1145/3411764.3445108}
\showDOI{\tempurl}


\bibitem[Awad et~al\mbox{.}(2018)]%
        {Awad2018}
\bibfield{author}{\bibinfo{person}{Edmond Awad}, \bibinfo{person}{Sohan Dsouza}, \bibinfo{person}{Richard Kim}, \bibinfo{person}{Jonathan Schulz}, \bibinfo{person}{Joseph Henrich}, \bibinfo{person}{Azim Shariff}, \bibinfo{person}{Jean-Fran{\c c}ois Bonnefon}, {and} \bibinfo{person}{Iyad Rahwan}.} \bibinfo{year}{2018}\natexlab{}.
\newblock \showarticletitle{The Moral Machine experiment}.
\newblock \bibinfo{journal}{\emph{Nature}} \bibinfo{volume}{563}, \bibinfo{number}{7729} (\bibinfo{date}{Nov.} \bibinfo{year}{2018}), \bibinfo{pages}{59--64}.
\newblock


\bibitem[Bauer and Schedl(2019)]%
        {Bauer2019}
\bibfield{author}{\bibinfo{person}{Christine Bauer} {and} \bibinfo{person}{Markus Schedl}.} \bibinfo{year}{2019}\natexlab{}.
\newblock \showarticletitle{Cross-country user connections in an online social network for music}.
\newblock \bibinfo{journal}{\emph{Extended Abstracts of the 2019 CHI Conference on Human Factors in Computing Systems}}.
\newblock


\bibitem[Baughan et~al\mbox{.}(2021)]%
        {Baughan2021}
\bibfield{author}{\bibinfo{person}{Amanda Baughan}, \bibinfo{person}{Nigini Oliveira}, \bibinfo{person}{Tal August}, \bibinfo{person}{Naomi Yamashita}, {and} \bibinfo{person}{Katharina Reinecke}.} \bibinfo{year}{2021}\natexlab{}.
\newblock \showarticletitle{Do cross-cultural differences in visual attention patterns affect search efficiency on websites?}
\newblock \bibinfo{journal}{\emph{Proceedings of the 2021 CHI Conference on Human Factors in Computing Systems}}.
\newblock
\urldef\tempurl%
\url{https://doi.org/10.1145/3411764.3445519}
\showDOI{\tempurl}


\bibitem[Baumeister and Leary(1995)]%
        {baumeister1995need}
\bibfield{author}{\bibinfo{person}{Roy~F. Baumeister} {and} \bibinfo{person}{Mark~R. Leary}.} \bibinfo{year}{1995}\natexlab{}.
\newblock \showarticletitle{The Need to Belong: Desire for Interpersonal Attachments as a Fundamental Human Motivation}.
\newblock \bibinfo{journal}{\emph{Psychological Bulletin}} \bibinfo{volume}{117}, \bibinfo{number}{3} (\bibinfo{year}{1995}), \bibinfo{pages}{497–529}.
\newblock
\urldef\tempurl%
\url{https://doi.org/10.1037/0033-2909.117.3.497}
\showDOI{\tempurl}


\bibitem[Boiger et~al\mbox{.}(2012)]%
        {Boiger2012}
\bibfield{author}{\bibinfo{person}{Michael Boiger}, \bibinfo{person}{Batja Mesquita}, \bibinfo{person}{Annie~Y. Tsai}, {and} \bibinfo{person}{Hazel Markus}.} \bibinfo{year}{2012}\natexlab{}.
\newblock \showarticletitle{Influencing and adjusting in daily emotional situations: A comparison of European and Asian American action styles}.
\newblock \bibinfo{journal}{\emph{Cognition and Emotion}} \bibinfo{volume}{26}, \bibinfo{number}{2} (\bibinfo{year}{2012}), \bibinfo{pages}{332--340}.
\newblock
\urldef\tempurl%
\url{https://doi.org/10.1080/02699931.2011.572422}
\showDOI{\tempurl}
\showeprint{https://doi.org/10.1080/02699931.2011.572422}
\newblock
\shownote{PMID: 21707271}.


\bibitem[Bostrom(2014)]%
        {Bostrom2014}
\bibfield{author}{\bibinfo{person}{Nick Bostrom}.} \bibinfo{year}{2014}\natexlab{}.
\newblock \bibinfo{booktitle}{\emph{Superintelligence: Paths, dangers, strategies}}.
\newblock \bibinfo{publisher}{Oxford University Press}.
\newblock


\bibitem[Brannon et~al\mbox{.}(2015)]%
        {Brannon2015}
\bibfield{author}{\bibinfo{person}{Tiffany~N Brannon}, \bibinfo{person}{Hazel~Rose Markus}, {and} \bibinfo{person}{Valerie~Jones Taylor}.} \bibinfo{year}{2015}\natexlab{}.
\newblock \showarticletitle{``Two souls, two thoughts,'' two self-schemas: double consciousness can have positive academic consequences for African Americans}.
\newblock \bibinfo{journal}{\emph{J. Pers. Soc. Psychol.}} \bibinfo{volume}{108}, \bibinfo{number}{4} (\bibinfo{year}{2015}), \bibinfo{pages}{586--609}.
\newblock
\urldef\tempurl%
\url{https://doi.org/10.1037/a0038992}
\showDOI{\tempurl}


\bibitem[Brynjolfsson and Mitchell(2017)]%
        {brynjolfsson2017machine}
\bibfield{author}{\bibinfo{person}{Erik Brynjolfsson} {and} \bibinfo{person}{Tom Mitchell}.} \bibinfo{year}{2017}\natexlab{}.
\newblock \showarticletitle{What Can Machine Learning Do? Workforce Implications}.
\newblock \bibinfo{journal}{\emph{Science}} \bibinfo{volume}{358}, \bibinfo{number}{6370} (\bibinfo{year}{2017}), \bibinfo{pages}{1530--1534}.
\newblock


\bibitem[Cave and Dihal(2019)]%
        {Cave2019}
\bibfield{author}{\bibinfo{person}{Stephen Cave} {and} \bibinfo{person}{Kanta Dihal}.} \bibinfo{year}{2019}\natexlab{}.
\newblock \showarticletitle{Hopes and fears for intelligent machines in fiction and reality}.
\newblock \bibinfo{journal}{\emph{Nature Machine Intelligence}}  \bibinfo{volume}{1} (\bibinfo{year}{2019}), \bibinfo{pages}{74--78}.
\newblock
Issue 2.
\showISSN{2522-5839}
\urldef\tempurl%
\url{https://doi.org/10.1038/s42256-019-0020-9}
\showDOI{\tempurl}


\bibitem[Chaves and Gerosa(2021)]%
        {Chaves2021}
\bibfield{author}{\bibinfo{person}{Ana~Paula Chaves} {and} \bibinfo{person}{Marco~Aurelio Gerosa}.} \bibinfo{year}{2021}\natexlab{}.
\newblock \showarticletitle{How should my chatbot interact? A survey on social characteristics in human–chatbot interaction design}.
\newblock \bibinfo{journal}{\emph{Int. J. Hum. Comput. Interact.}}  \bibinfo{volume}{37} (\bibinfo{date}{5} \bibinfo{year}{2021}), \bibinfo{pages}{729--758}.
\newblock
Issue 8.


\bibitem[Cheng et~al\mbox{.}(2021)]%
        {Cheng2021}
\bibfield{author}{\bibinfo{person}{Justin Cheng}, \bibinfo{person}{Moira Burke}, {and} \bibinfo{person}{Bethany de Gant}.} \bibinfo{year}{2021}\natexlab{}.
\newblock \showarticletitle{Country Differences in Social Comparison on Social Media}.
\newblock \bibinfo{journal}{\emph{Proceedings of the ACM on Human-Computer Interaction}}  \bibinfo{volume}{4} (\bibinfo{date}{1} \bibinfo{year}{2021}), \bibinfo{pages}{1--26}.
\newblock
Issue CSCW3.
\showISSN{2573-0142}
\urldef\tempurl%
\url{https://doi.org/10.1145/3434179}
\showDOI{\tempurl}


\bibitem[Chowdhury and Wynn(2011)]%
        {Chowdhury2011}
\bibfield{author}{\bibinfo{person}{Sujoy~Kumar Chowdhury} {and} \bibinfo{person}{Jody Wynn}.} \bibinfo{year}{2011}\natexlab{}.
\newblock \showarticletitle{Cowabunga! A System to Facilitate Multi-Cultural Diversity through Couchsurfing}.
\newblock \bibinfo{journal}{\emph{CHI '11 Extended Abstracts on Human Factors in Computing Systems}}, \bibinfo{pages}{965–970}.
\newblock
\showISBNx{9781450302685}
\urldef\tempurl%
\url{https://doi.org/10.1145/1979742.1979506}
\showDOI{\tempurl}


\bibitem[Cohen and Varnum(2016)]%
        {Cohen2016}
\bibfield{author}{\bibinfo{person}{Adam~B Cohen} {and} \bibinfo{person}{Michael~EW Varnum}.} \bibinfo{year}{2016}\natexlab{}.
\newblock \showarticletitle{Beyond East vs. West: social class, region, and religion as forms of culture}.
\newblock \bibinfo{journal}{\emph{Current Opinion in Psychology}}  \bibinfo{volume}{8} (\bibinfo{date}{4} \bibinfo{year}{2016}), \bibinfo{pages}{5--9}.
\newblock
\showISSN{2352250X}
\urldef\tempurl%
\url{https://doi.org/10.1016/j.copsyc.2015.09.006}
\showDOI{\tempurl}


\bibitem[Comber et~al\mbox{.}(2020)]%
        {Comber2020}
\bibfield{author}{\bibinfo{person}{Rob Comber}, \bibinfo{person}{Shaowen Bardzell}, \bibinfo{person}{Jeffrey Bardzell}, \bibinfo{person}{Mike Hazas}, {and} \bibinfo{person}{Michael Muller}.} \bibinfo{year}{2020}\natexlab{}.
\newblock \showarticletitle{Announcing a new CHI subcommittee}.
\newblock \bibinfo{journal}{\emph{Interactions}}  \bibinfo{volume}{27} (\bibinfo{date}{7} \bibinfo{year}{2020}), \bibinfo{pages}{101--103}.
\newblock
Issue 4.
\showISSN{1072-5520}
\urldef\tempurl%
\url{https://doi.org/10.1145/3407228}
\showDOI{\tempurl}


\bibitem[Debruge(2015)]%
        {Debruge15}
\bibfield{author}{\bibinfo{person}{Peter Debruge}.} \bibinfo{year}{2015}\natexlab{}.
\newblock \bibinfo{title}{Film Review: ‘Sayonara’}.
\newblock \bibinfo{howpublished}{\url{https://variety.com/2015/film/reviews/sayonara-film-review-1201625975}}.
\newblock
\newblock
\shownote{Accessed: 2023-9-15}.


\bibitem[Devendorf and Goodman(2014)]%
        {Devendorf2014}
\bibfield{author}{\bibinfo{person}{Laura Devendorf} {and} \bibinfo{person}{Elizabeth Goodman}.} \bibinfo{year}{2014}\natexlab{}.
\newblock \bibinfo{title}{The Algorithm Multiple, the Algorithm Material: Reconstructing Creative Practice}.
\newblock
\newblock
\newblock
\shownote{Talk given at The Contours of Algorithmic Life, a conference at UC Davis -- May 15-16, 2014.}.


\bibitem[Doctor et~al\mbox{.}(2022)]%
        {Doctor2022}
\bibfield{author}{\bibinfo{person}{Thomas Doctor}, \bibinfo{person}{Olaf Witkowski}, \bibinfo{person}{Elizaveta Solomonova}, \bibinfo{person}{Bill Duane}, {and} \bibinfo{person}{Michael Levin}.} \bibinfo{year}{2022}\natexlab{}.
\newblock \showarticletitle{Biology, Buddhism, and AI: Care as the Driver of Intelligence}.
\newblock \bibinfo{journal}{\emph{Entropy}}  \bibinfo{volume}{24} (\bibinfo{date}{5} \bibinfo{year}{2022}), \bibinfo{pages}{710}.
\newblock
Issue 5.
\showISSN{1099-4300}
\urldef\tempurl%
\url{https://doi.org/10.3390/e24050710}
\showDOI{\tempurl}


\bibitem[Druin(2018)]%
        {Druin2018}
\bibfield{author}{\bibinfo{person}{Allison Druin}.} \bibinfo{year}{2018}\natexlab{}.
\newblock \bibinfo{title}{The Possibilities of Inclusion for SIGCHI}.
\newblock
\newblock
\newblock
\shownote{From https://sigchi.org/the-possibilities-of-inclusion-for-sigchi/}.


\bibitem[Dutta(2011)]%
        {Dutta2011}
\bibfield{author}{\bibinfo{person}{Mohan~J Dutta}.} \bibinfo{year}{2011}\natexlab{}.
\newblock \bibinfo{booktitle}{\emph{Communicating Social Change}}.
\newblock \bibinfo{publisher}{Routledge}, \bibinfo{address}{London, England}.
\newblock


\bibitem[Emile H.L.~Aarts(2006)]%
        {Aarts2006}
\bibfield{author}{\bibinfo{person}{José Luis~Encarna{\c c}{\~a}o Emile H.L.~Aarts}.} \bibinfo{year}{2006}\natexlab{}.
\newblock \bibinfo{booktitle}{\emph{True Visions: The Emergence of Ambient Intelligence}}.
\newblock \bibinfo{publisher}{Springer}, \bibinfo{address}{Germany; Berlin Heidelberg}.
\newblock


\bibitem[Epley et~al\mbox{.}(2007)]%
        {Epley2007}
\bibfield{author}{\bibinfo{person}{Nicholas Epley}, \bibinfo{person}{Adam Waytz}, {and} \bibinfo{person}{John~T. Cacioppo}.} \bibinfo{year}{2007}\natexlab{}.
\newblock \showarticletitle{On seeing human: A three-factor theory of anthropomorphism.}
\newblock \bibinfo{journal}{\emph{Psychological Review}}  \bibinfo{volume}{114} (\bibinfo{year}{2007}), \bibinfo{pages}{864--886}.
\newblock
Issue 4.
\showISSN{1939-1471}
\urldef\tempurl%
\url{https://doi.org/10.1037/0033-295X.114.4.864}
\showDOI{\tempurl}


\bibitem[Evers et~al\mbox{.}(2008)]%
        {Evers2008}
\bibfield{author}{\bibinfo{person}{Vanessa Evers}, \bibinfo{person}{Heidy~C. Maldonado}, \bibinfo{person}{Talia~L. Brodecki}, {and} \bibinfo{person}{Pamela~J. Hinds}.} \bibinfo{year}{2008}\natexlab{}.
\newblock \showarticletitle{Relational vs. Group Self-Construal: Untangling the Role of National Culture in HRI}. In \bibinfo{booktitle}{\emph{Proceedings of the 3rd ACM/IEEE International Conference on Human Robot Interaction}} (Amsterdam, The Netherlands) \emph{(\bibinfo{series}{HRI '08})}. \bibinfo{publisher}{Association for Computing Machinery}, \bibinfo{address}{New York, NY, USA}, \bibinfo{pages}{255–262}.
\newblock
\showISBNx{9781605580173}
\urldef\tempurl%
\url{https://doi.org/10.1145/1349822.1349856}
\showDOI{\tempurl}


\bibitem[Fiske et~al\mbox{.}(1998)]%
        {Fiske1998}
\bibfield{author}{\bibinfo{person}{Alan~Page Fiske}, \bibinfo{person}{Shinobu Kitayama}, \bibinfo{person}{Hazel~Rose Markus}, {and} \bibinfo{person}{Richard~E Nisbett}.} \bibinfo{year}{1998}\natexlab{}.
\newblock \showarticletitle{The cultural matrix of social psychology.}
\newblock In \bibinfo{booktitle}{\emph{The handbook of social psychology, Vols. 1-2, 4th ed.}} \bibinfo{publisher}{McGraw-Hill}, \bibinfo{pages}{915--981}.
\newblock
\showISBNx{0-19-521376-9 (Hardcover)}


\bibitem[Friedman et~al\mbox{.}(2003)]%
        {Friedman2003}
\bibfield{author}{\bibinfo{person}{Batya Friedman}, \bibinfo{person}{Peter~H. Kahn}, {and} \bibinfo{person}{Jennifer Hagman}.} \bibinfo{year}{2003}\natexlab{}.
\newblock \showarticletitle{Hardware Companions? What Online AIBO Discussion Forums Reveal about the Human-Robotic Relationship}. In \bibinfo{booktitle}{\emph{Proceedings of the SIGCHI Conference on Human Factors in Computing Systems}} (Ft. Lauderdale, Florida, USA) \emph{(\bibinfo{series}{CHI '03})}. \bibinfo{publisher}{Association for Computing Machinery}, \bibinfo{address}{New York, NY, USA}, \bibinfo{pages}{273–280}.
\newblock
\showISBNx{1581136307}
\urldef\tempurl%
\url{https://doi.org/10.1145/642611.642660}
\showDOI{\tempurl}


\bibitem[Friend(2018)]%
        {Friend2018}
\bibfield{author}{\bibinfo{person}{Tad Friend}.} \bibinfo{year}{2018}\natexlab{}.
\newblock \showarticletitle{How frightened should we be of A.I.?}
\newblock \bibinfo{journal}{\emph{The New Yorker. Retrieved September 7, 2022, from https://www.newyorker.com/magazine/2018/05/14/how-frightened-should-we-be-of-ai}} (\bibinfo{date}{5} \bibinfo{year}{2018}).
\newblock


\bibitem[Furnham and Ribchester(1995)]%
        {furnham1995tolerance}
\bibfield{author}{\bibinfo{person}{Adrian Furnham} {and} \bibinfo{person}{Trevor Ribchester}.} \bibinfo{year}{1995}\natexlab{}.
\newblock \showarticletitle{Tolerance of Ambiguity: A Review of the Concept, Its Measurement and Applications}.
\newblock \bibinfo{journal}{\emph{Current Psychology}}  \bibinfo{volume}{14} (\bibinfo{year}{1995}), \bibinfo{pages}{179–199}.
\newblock
\urldef\tempurl%
\url{https://doi.org/10.1007/BF02686907}
\showDOI{\tempurl}


\bibitem[GAIN(nd)]%
        {GAIN}
\bibfield{author}{\bibinfo{person}{GAIN}.} \bibinfo{year}{n.d.}\natexlab{}.
\newblock \bibinfo{title}{Global AI narratives}.
\newblock
\newblock
\newblock
\shownote{Retrieved September 7, 2022, from https://www.ainarratives.com/}.


\bibitem[Gelfand and Kashima(2016)]%
        {Gelfand2016}
\bibfield{author}{\bibinfo{person}{Michele~J Gelfand} {and} \bibinfo{person}{Yoshihisa Kashima}.} \bibinfo{year}{2016}\natexlab{}.
\newblock \showarticletitle{Editorial overview: Culture: Advances in the science of culture and psychology}.
\newblock \bibinfo{journal}{\emph{Current Opinion in Psychology}}  \bibinfo{volume}{8} (\bibinfo{date}{4} \bibinfo{year}{2016}), \bibinfo{pages}{iv--x}.
\newblock
\showISSN{2352250X}
\urldef\tempurl%
\url{https://doi.org/10.1016/j.copsyc.2015.12.011}
\showDOI{\tempurl}


\bibitem[Gray et~al\mbox{.}(2007)]%
        {Gray2007}
\bibfield{author}{\bibinfo{person}{Heather~M. Gray}, \bibinfo{person}{Kurt Gray}, {and} \bibinfo{person}{Daniel~M. Wegner}.} \bibinfo{year}{2007}\natexlab{}.
\newblock \showarticletitle{Dimensions of Mind Perception}.
\newblock \bibinfo{journal}{\emph{Science}} \bibinfo{volume}{315}, \bibinfo{number}{5812} (\bibinfo{year}{2007}), \bibinfo{pages}{619--619}.
\newblock
\urldef\tempurl%
\url{https://doi.org/10.1126/science.1134475}
\showDOI{\tempurl}
\showeprint{https://www.science.org/doi/pdf/10.1126/science.1134475}


\bibitem[Guan et~al\mbox{.}(2015)]%
        {Guan2015}
\bibfield{author}{\bibinfo{person}{Yanjun Guan}, \bibinfo{person}{Sylvia~Xiaohua Chen}, \bibinfo{person}{Nimrod Levin}, \bibinfo{person}{Michael~Harris Bond}, \bibinfo{person}{Nanfeng Luo}, \bibinfo{person}{Jingwen Xu}, \bibinfo{person}{Xiang Zhou}, \bibinfo{person}{Pei Chen}, \bibinfo{person}{Chendi Li}, \bibinfo{person}{Ruchunyi Fu}, \bibinfo{person}{Jiawei Zhang}, \bibinfo{person}{Yueting Ji}, \bibinfo{person}{Zichuan Mo}, \bibinfo{person}{Yumeng Li}, \bibinfo{person}{Zheng Fang}, \bibinfo{person}{Dongqian Jiang}, {and} \bibinfo{person}{Xue Han}.} \bibinfo{year}{2015}\natexlab{}.
\newblock \showarticletitle{Differences in Career Decision-Making Profiles Between American and Chinese University Students: The Relative Strength of Mediating Mechanisms Across Cultures}.
\newblock \bibinfo{journal}{\emph{Journal of Cross-Cultural Psychology}} \bibinfo{volume}{46}, \bibinfo{number}{6} (\bibinfo{year}{2015}), \bibinfo{pages}{856--872}.
\newblock
\urldef\tempurl%
\url{https://doi.org/10.1177/0022022115585874}
\showDOI{\tempurl}
\showeprint{https://doi.org/10.1177/0022022115585874}


\bibitem[Gutierrez and Ochoa(2016)]%
        {Gutierrez2016}
\bibfield{author}{\bibinfo{person}{Francisco~J. Gutierrez} {and} \bibinfo{person}{Sergio~F. Ochoa}.} \bibinfo{year}{2016}\natexlab{}.
\newblock \showarticletitle{Mom, I Do Have a Family!}
\newblock \bibinfo{journal}{\emph{Proceedings of the 19th ACM Conference on Computer-Supported Cooperative Work \& Social Computing}}, \bibinfo{pages}{1402--1411}.
\newblock
\showISBNx{9781450335928}
\urldef\tempurl%
\url{https://doi.org/10.1145/2818048.2820000}
\showDOI{\tempurl}


\bibitem[Haddad et~al\mbox{.}(2014)]%
        {Haddad2014}
\bibfield{author}{\bibinfo{person}{Shathel Haddad}, \bibinfo{person}{Joanna McGrenere}, {and} \bibinfo{person}{Claudia Jacova}.} \bibinfo{year}{2014}\natexlab{}.
\newblock \showarticletitle{Interface Design for Older Adults with Varying Cultural Attitudes toward Uncertainty}. In \bibinfo{booktitle}{\emph{Proceedings of the SIGCHI Conference on Human Factors in Computing Systems}} (Toronto, Ontario, Canada) \emph{(\bibinfo{series}{CHI '14})}. \bibinfo{publisher}{Association for Computing Machinery}, \bibinfo{address}{New York, NY, USA}, \bibinfo{pages}{1913–1922}.
\newblock
\showISBNx{9781450324731}
\urldef\tempurl%
\url{https://doi.org/10.1145/2556288.2557124}
\showDOI{\tempurl}


\bibitem[Han and Humphreys(2016)]%
        {Han2016}
\bibfield{author}{\bibinfo{person}{Shihui Han} {and} \bibinfo{person}{Glyn Humphreys}.} \bibinfo{year}{2016}\natexlab{}.
\newblock \showarticletitle{Self-construal: a cultural framework for brain function}.
\newblock \bibinfo{journal}{\emph{Current Opinion in Psychology}}  \bibinfo{volume}{8} (\bibinfo{year}{2016}), \bibinfo{pages}{10--14}.
\newblock
\showISSN{2352-250X}
\urldef\tempurl%
\url{https://doi.org/10.1016/j.copsyc.2015.09.013}
\showDOI{\tempurl}
\newblock
\shownote{Culture}.


\bibitem[Haraway(2006)]%
        {Haraway2006}
\bibfield{author}{\bibinfo{person}{Donna Haraway}.} \bibinfo{year}{2006}\natexlab{}.
\newblock \bibinfo{booktitle}{\emph{A Cyborg Manifesto: Science, Technology, and Socialist-Feminism in the Late 20th Century}}.
\newblock \bibinfo{publisher}{Springer Netherlands}, \bibinfo{address}{Dordrecht}, \bibinfo{pages}{117--158}.
\newblock
\showISBNx{978-1-4020-3803-7}
\urldef\tempurl%
\url{https://doi.org/10.1007/978-1-4020-3803-7_4}
\showDOI{\tempurl}


\bibitem[Hazel~Rose and Conner(2014)]%
        {Markus2014}
\bibfield{author}{\bibinfo{person}{Markus Hazel~Rose} {and} \bibinfo{person}{Alana Conner}.} \bibinfo{year}{2014}\natexlab{}.
\newblock \bibinfo{booktitle}{\emph{Clash!: How to thrive in a multicultural world}}.
\newblock \bibinfo{publisher}{Penguin}.
\newblock
\showISBNx{9780142180938}


\bibitem[He et~al\mbox{.}(2010)]%
        {He2010}
\bibfield{author}{\bibinfo{person}{Yurong He}, \bibinfo{person}{Chen Zhao}, {and} \bibinfo{person}{Pamela Hinds}.} \bibinfo{year}{2010}\natexlab{}.
\newblock \showarticletitle{Understanding information sharing from a cross-cultural perspective}. In \bibinfo{booktitle}{\emph{CHI '10 Extended Abstracts on Human Factors in Computing Systems}} (Atlanta, Georgia, USA) \emph{(\bibinfo{series}{CHI EA '10})}. \bibinfo{publisher}{Association for Computing Machinery}, \bibinfo{address}{New York, NY, USA}, \bibinfo{pages}{3823–3828}.
\newblock
\showISBNx{9781605589305}
\urldef\tempurl%
\url{https://doi.org/10.1145/1753846.1754063}
\showDOI{\tempurl}


\bibitem[Heimgärtner(2013)]%
        {Heimgartner2013}
\bibfield{author}{\bibinfo{person}{Rüdiger Heimgärtner}.} \bibinfo{year}{2013}\natexlab{}.
\newblock \showarticletitle{Reflections on a Model of Culturally Influenced Human–Computer Interaction to Cover Cultural Contexts in HCI Design}.
\newblock \bibinfo{journal}{\emph{International Journal of Human–Computer Interaction}} \bibinfo{volume}{29}, \bibinfo{number}{4} (\bibinfo{year}{2013}), \bibinfo{pages}{205--219}.
\newblock
\urldef\tempurl%
\url{https://doi.org/10.1080/10447318.2013.765761}
\showDOI{\tempurl}
\showeprint{https://doi.org/10.1080/10447318.2013.765761}


\bibitem[Hekler et~al\mbox{.}(2013)]%
        {Hekler2013}
\bibfield{author}{\bibinfo{person}{Eric~B. Hekler}, \bibinfo{person}{Predrag Klasnja}, \bibinfo{person}{Jon~E. Froehlich}, {and} \bibinfo{person}{Matthew~P. Buman}.} \bibinfo{year}{2013}\natexlab{}.
\newblock \showarticletitle{Mind the theoretical gap: interpreting, using, and developing behavioral theory in HCI research}. In \bibinfo{booktitle}{\emph{Proceedings of the SIGCHI Conference on Human Factors in Computing Systems}} (Paris, France) \emph{(\bibinfo{series}{CHI '13})}. \bibinfo{publisher}{Association for Computing Machinery}, \bibinfo{address}{New York, NY, USA}, \bibinfo{pages}{3307–3316}.
\newblock
\showISBNx{9781450318990}
\urldef\tempurl%
\url{https://doi.org/10.1145/2470654.2466452}
\showDOI{\tempurl}


\bibitem[Henrich et~al\mbox{.}(2010)]%
        {Henrich2010}
\bibfield{author}{\bibinfo{person}{Joseph Henrich}, \bibinfo{person}{Steven~J. Heine}, {and} \bibinfo{person}{Ara Norenzayan}.} \bibinfo{year}{2010}\natexlab{}.
\newblock \showarticletitle{Most people are not WEIRD}.
\newblock \bibinfo{journal}{\emph{Nature}}  \bibinfo{volume}{466} (\bibinfo{date}{7} \bibinfo{year}{2010}), \bibinfo{pages}{29--29}.
\newblock
Issue 7302.
\showISSN{0028-0836}
\urldef\tempurl%
\url{https://doi.org/10.1038/466029a}
\showDOI{\tempurl}


\bibitem[Himmelsbach et~al\mbox{.}(2019)]%
        {Himmelsbach2019}
\bibfield{author}{\bibinfo{person}{Julia Himmelsbach}, \bibinfo{person}{Stephanie Schwarz}, \bibinfo{person}{Cornelia Gerdenitsch}, \bibinfo{person}{Beatrix Wais-Zechmann}, \bibinfo{person}{Jan Bobeth}, {and} \bibinfo{person}{Manfred Tscheligi}.} \bibinfo{year}{2019}\natexlab{}.
\newblock \showarticletitle{Do We Care About Diversity in Human Computer Interaction}.
\newblock \bibinfo{journal}{\emph{Proceedings of the 2019 CHI Conference on Human Factors in Computing Systems}}, \bibinfo{pages}{1--16}.
\newblock
\showISBNx{9781450359702}
\urldef\tempurl%
\url{https://doi.org/10.1145/3290605.3300720}
\showDOI{\tempurl}


\bibitem[Hofstede(1984)]%
        {Hofstede1984}
\bibfield{author}{\bibinfo{person}{Geert Hofstede}.} \bibinfo{year}{1984}\natexlab{}.
\newblock \bibinfo{booktitle}{\emph{Culture's consequences: Culture’s consequences:International differences in work-related values}}. Vol.~\bibinfo{volume}{5}.
\newblock \bibinfo{publisher}{Sage}.
\newblock


\bibitem[Hofstede(2001)]%
        {Hofstede2001}
\bibfield{author}{\bibinfo{person}{Geert Hofstede}.} \bibinfo{year}{2001}\natexlab{}.
\newblock \bibinfo{booktitle}{\emph{Culture's consequences: Comparing values, behaviors, institutions, and organizations across nations}}.
\newblock \bibinfo{publisher}{Sage}.
\newblock


\bibitem[Hwang et~al\mbox{.}(2018)]%
        {Hwang2018}
\bibfield{author}{\bibinfo{person}{Euijin Hwang}, \bibinfo{person}{Reuben Kirkham}, \bibinfo{person}{Andrew Monk}, {and} \bibinfo{person}{Patrick Olivier}.} \bibinfo{year}{2018}\natexlab{}.
\newblock \showarticletitle{Respectful Disconnection}.
\newblock \bibinfo{journal}{\emph{Proceedings of the 2018 Designing Interactive Systems Conference}}, \bibinfo{pages}{733--745}.
\newblock
\showISBNx{9781450351980}
\urldef\tempurl%
\url{https://doi.org/10.1145/3196709.3196751}
\showDOI{\tempurl}


\bibitem[Isbister(2019)]%
        {Isbister2019}
\bibfield{author}{\bibinfo{person}{Katherine Isbister}.} \bibinfo{year}{2019}\natexlab{}.
\newblock \showarticletitle{Toward ‘Suprahuman’ Technology}.
\newblock \bibinfo{journal}{\emph{Proceedings of the Halfway to the Future Symposium 2019}}, \bibinfo{pages}{1--4}.
\newblock
\showISBNx{9781450372039}
\urldef\tempurl%
\url{https://doi.org/10.1145/3363384.3363468}
\showDOI{\tempurl}


\bibitem[Ishiguro(2015)]%
        {Ishiguro2015}
\bibfield{author}{\bibinfo{person}{Hiroshi Ishiguro}.} \bibinfo{year}{2015}\natexlab{}.
\newblock \bibinfo{title}{Greetings from the stage at the 28th Tokyo International Film Festival (Sayonara)}.
\newblock
\newblock
\newblock
\shownote{Retrieved from https://youtu.be/nOEr3kzilXE}.


\bibitem[Jensen and Blok(2013)]%
        {Jensen2013}
\bibfield{author}{\bibinfo{person}{Casper~Bruun Jensen} {and} \bibinfo{person}{Anders Blok}.} \bibinfo{year}{2013}\natexlab{}.
\newblock \showarticletitle{Techno-animism in japan: Shinto cosmograms, actor-network theory, and the enabling powers of non-human agencies}.
\newblock \bibinfo{journal}{\emph{Theory Cult. Soc.}} \bibinfo{volume}{30}, \bibinfo{number}{2} (\bibinfo{date}{March} \bibinfo{year}{2013}), \bibinfo{pages}{84--115}.
\newblock


\bibitem[Kamppuri et~al\mbox{.}(2006)]%
        {Kamppuri2006}
\bibfield{author}{\bibinfo{person}{Minna Kamppuri}, \bibinfo{person}{Roman Bednarik}, {and} \bibinfo{person}{Markku Tukiainen}.} \bibinfo{year}{2006}\natexlab{}.
\newblock \showarticletitle{The expanding focus of HCI}.
\newblock \bibinfo{journal}{\emph{Proceedings of the 4th Nordic conference on Human-computer interaction changing roles - NordiCHI '06}}, \bibinfo{pages}{405--408}.
\newblock
\showISBNx{1595933255}
\urldef\tempurl%
\url{https://doi.org/10.1145/1182475.1182523}
\showDOI{\tempurl}


\bibitem[Kim et~al\mbox{.}(2010)]%
        {Kim2010}
\bibfield{author}{\bibinfo{person}{Jang~Hyun Kim}, \bibinfo{person}{Min-Sun Kim}, {and} \bibinfo{person}{Yoonjae Nam}.} \bibinfo{year}{2010}\natexlab{}.
\newblock \showarticletitle{An Analysis of Self-Construals, Motivations, Facebook Use, and User Satisfaction}.
\newblock \bibinfo{journal}{\emph{International Journal of Human-Computer Interaction}}  \bibinfo{volume}{26} (\bibinfo{date}{11} \bibinfo{year}{2010}), \bibinfo{pages}{1077--1099}.
\newblock
Issue 11-12.
\showISSN{1044-7318}
\urldef\tempurl%
\url{https://doi.org/10.1080/10447318.2010.516726}
\showDOI{\tempurl}


\bibitem[Kistler et~al\mbox{.}(2012)]%
        {Kistler2012}
\bibfield{author}{\bibinfo{person}{Felix Kistler}, \bibinfo{person}{Birgit Endrass}, \bibinfo{person}{Ionut Damian}, \bibinfo{person}{Chi~Tai Dang}, {and} \bibinfo{person}{Elisabeth André}.} \bibinfo{year}{2012}\natexlab{}.
\newblock \showarticletitle{Natural interaction with culturally adaptive virtual characters}.
\newblock \bibinfo{journal}{\emph{Journal on Multimodal User Interfaces}}  \bibinfo{volume}{6} (\bibinfo{date}{7} \bibinfo{year}{2012}), \bibinfo{pages}{39--47}.
\newblock
Issue 1-2.
\showISSN{1783-7677}
\urldef\tempurl%
\url{https://doi.org/10.1007/s12193-011-0087-z}
\showDOI{\tempurl}


\bibitem[Kitano(2007)]%
        {Kitano2007}
\bibfield{author}{\bibinfo{person}{Naho Kitano}.} \bibinfo{year}{2007}\natexlab{}.
\newblock \showarticletitle{Animism , Rinri , Modernization ; the Base of Japanese Robotics}.
\newblock


\bibitem[Kitayama et~al\mbox{.}(2009)]%
        {Kitayama2009-eu}
\bibfield{author}{\bibinfo{person}{Shinobu Kitayama}, \bibinfo{person}{Hyekyung Park}, \bibinfo{person}{A~Timur Sevincer}, \bibinfo{person}{Mayumi Karasawa}, {and} \bibinfo{person}{Ayse~K Uskul}.} \bibinfo{year}{2009}\natexlab{}.
\newblock \showarticletitle{A cultural task analysis of implicit independence: comparing North America, Western Europe, and East Asia}.
\newblock \bibinfo{journal}{\emph{J. Pers. Soc. Psychol.}} \bibinfo{volume}{97}, \bibinfo{number}{2} (\bibinfo{date}{Aug.} \bibinfo{year}{2009}), \bibinfo{pages}{236--255}.
\newblock


\bibitem[Kitayama and Uchida(2005)]%
        {Kitayama2005}
\bibfield{author}{\bibinfo{person}{Shinobu Kitayama} {and} \bibinfo{person}{Yuchida Uchida}.} \bibinfo{year}{2005}\natexlab{}.
\newblock \showarticletitle{Interdependent agency: An alternative system for action}.
\newblock \bibinfo{journal}{\emph{Cultural and social behavior: The Ontario symposium}}  \bibinfo{volume}{10} (\bibinfo{year}{2005}), \bibinfo{pages}{137--164}.
\newblock


\bibitem[Kopecka and Such(2020)]%
        {Kopecka2020}
\bibfield{author}{\bibinfo{person}{Hana Kopecka} {and} \bibinfo{person}{Jose~M Such}.} \bibinfo{year}{2020}\natexlab{}.
\newblock \showarticletitle{Explainable AI for Cultural Minds}.
\newblock \bibinfo{journal}{\emph{Workshop on Dialogue, Explanation and Argumentation for Human-Agent Interaction, DEXAHAI}}.
\newblock
\urldef\tempurl%
\url{https://sites.google.com/view/dexahai-at-ecai2020/home}
\showURL{%
\tempurl}
\newblock
\shownote{Conference date: 07-09-2020}.


\bibitem[Korn et~al\mbox{.}(2021)]%
        {Korn2021}
\bibfield{author}{\bibinfo{person}{Oliver Korn}, \bibinfo{person}{Neziha Akalin}, {and} \bibinfo{person}{Ruben Gouveia}.} \bibinfo{year}{2021}\natexlab{}.
\newblock \showarticletitle{Understanding Cultural Preferences for Social Robots: A Study in German and Arab Communities}.
\newblock \bibinfo{journal}{\emph{J. Hum.-Robot Interact.}} \bibinfo{volume}{10}, \bibinfo{number}{2}, Article \bibinfo{articleno}{12} (\bibinfo{date}{mar} \bibinfo{year}{2021}), \bibinfo{numpages}{19}~pages.
\newblock
\urldef\tempurl%
\url{https://doi.org/10.1145/3439717}
\showDOI{\tempurl}


\bibitem[Kumar et~al\mbox{.}(2018)]%
        {Kumar2018}
\bibfield{author}{\bibinfo{person}{Neha Kumar}, \bibinfo{person}{Kurtis Heimerl}, \bibinfo{person}{David Nemer}, \bibinfo{person}{Naveena Karusala}, \bibinfo{person}{Aditya Vashistha}, \bibinfo{person}{Susan~M. Dray}, \bibinfo{person}{Christian Sturm}, \bibinfo{person}{Laura~S. Gaytán-Lugo}, \bibinfo{person}{Anicia Peters}, \bibinfo{person}{Nova Ahmed}, \bibinfo{person}{Nicola Dell}, {and} \bibinfo{person}{Jay Chen}.} \bibinfo{year}{2018}\natexlab{}.
\newblock \showarticletitle{HCI Across Borders}.
\newblock \bibinfo{journal}{\emph{Extended Abstracts of the 2018 CHI Conference on Human Factors in Computing Systems}}, \bibinfo{pages}{1--8}.
\newblock
\showISBNx{9781450356213}
\urldef\tempurl%
\url{https://doi.org/10.1145/3170427.3170666}
\showDOI{\tempurl}


\bibitem[Lee and Sabanović(2014)]%
        {Lee2014}
\bibfield{author}{\bibinfo{person}{Hee~Rin Lee} {and} \bibinfo{person}{Selma Sabanović}.} \bibinfo{year}{2014}\natexlab{}.
\newblock \showarticletitle{Culturally variable preferences for robot design and use in South Korea, Turkey, and the United States}.
\newblock \bibinfo{journal}{\emph{Proceedings of the 2014 ACM/IEEE international conference on Human-robot interaction}}.
\newblock


\bibitem[Lee et~al\mbox{.}(2012)]%
        {Lee2012}
\bibfield{author}{\bibinfo{person}{Hee~Rin Lee}, \bibinfo{person}{JaYoung Sung}, \bibinfo{person}{Selma Šabanović}, {and} \bibinfo{person}{Joenghye Han}.} \bibinfo{year}{2012}\natexlab{}.
\newblock \showarticletitle{Cultural design of domestic robots: A study of user expectations in Korea and the United States}. In \bibinfo{booktitle}{\emph{2012 IEEE RO-MAN: The 21st IEEE International Symposium on Robot and Human Interactive Communication}}. \bibinfo{pages}{803--808}.
\newblock
\urldef\tempurl%
\url{https://doi.org/10.1109/ROMAN.2012.6343850}
\showDOI{\tempurl}


\bibitem[Lee et~al\mbox{.}(2015)]%
        {Lee2015}
\bibfield{author}{\bibinfo{person}{Min~Kyung Lee}, \bibinfo{person}{Nathaniel Fruchter}, {and} \bibinfo{person}{Laura Dabbish}.} \bibinfo{year}{2015}\natexlab{}.
\newblock \showarticletitle{Making Decisions From a Distance}.
\newblock \bibinfo{journal}{\emph{Proceedings of the 18th ACM Conference on Computer Supported Cooperative Work \& Social Computing}}, \bibinfo{pages}{1576--1589}.
\newblock
\showISBNx{9781450329224}
\urldef\tempurl%
\url{https://doi.org/10.1145/2675133.2675288}
\showDOI{\tempurl}


\bibitem[Lee and Rich(2021)]%
        {Lee2021}
\bibfield{author}{\bibinfo{person}{Min~Kyung Lee} {and} \bibinfo{person}{Katherine Rich}.} \bibinfo{year}{2021}\natexlab{}.
\newblock \showarticletitle{Who is included in human perceptions of AI?: Trust and perceived fairness around healthcare AI and cultural mistrust}.
\newblock \bibinfo{journal}{\emph{Proceedings of the 2021 CHI Conference on Human Factors in Computing Systems}}.
\newblock


\bibitem[Lewin(1999)]%
        {Lewin1999}
\bibfield{author}{\bibinfo{person}{Kurt~Z. Lewin}.} \bibinfo{year}{1999}\natexlab{}.
\newblock \showarticletitle{Intention, will and need (1st ed. 1926)}.
\newblock In \bibinfo{booktitle}{\emph{A Kurt Lewin Reader. The Complete Social Scientist}}, \bibfield{editor}{\bibinfo{person}{M~Gold}} (Ed.). \bibinfo{publisher}{American Psychological Association}, \bibinfo{pages}{83 – 115}.
\newblock


\bibitem[Lewis et~al\mbox{.}(2018)]%
        {Lewis2018}
\bibfield{author}{\bibinfo{person}{Jason~Edward Lewis}, \bibinfo{person}{Noelani Arista}, \bibinfo{person}{Archer Pechawis}, {and} \bibinfo{person}{Suzanne Kite}.} \bibinfo{year}{2018}\natexlab{}.
\newblock \showarticletitle{Making Kin with the Machines}.
\newblock \bibinfo{journal}{\emph{Journal of Design and Science}} (\bibinfo{date}{7} \bibinfo{year}{2018}).
\newblock
\urldef\tempurl%
\url{https://doi.org/10.21428/bfafd97b}
\showDOI{\tempurl}


\bibitem[Li et~al\mbox{.}(2010)]%
        {Li2010}
\bibfield{author}{\bibinfo{person}{Dingjun Li}, \bibinfo{person}{P.~L.~Patrick Rau}, {and} \bibinfo{person}{Ye Li}.} \bibinfo{year}{2010}\natexlab{}.
\newblock \showarticletitle{A Cross-cultural Study: Effect of Robot Appearance and Task}.
\newblock \bibinfo{journal}{\emph{International Journal of Social Robotics}}  \bibinfo{volume}{2} (\bibinfo{date}{6} \bibinfo{year}{2010}), \bibinfo{pages}{175--186}.
\newblock
Issue 2.
\showISSN{1875-4791}
\urldef\tempurl%
\url{https://doi.org/10.1007/s12369-010-0056-9}
\showDOI{\tempurl}


\bibitem[Linxen et~al\mbox{.}(2021a)]%
        {Linxen2021_review}
\bibfield{author}{\bibinfo{person}{Sebastian Linxen}, \bibinfo{person}{Vincent Cassau}, {and} \bibinfo{person}{Christian Sturm}.} \bibinfo{year}{2021}\natexlab{a}.
\newblock \showarticletitle{Culture and HCI: A still slowly growing field of research. Findings from a systematic, comparative mapping review}. In \bibinfo{booktitle}{\emph{Proceedings of the XXI International Conference on Human Computer Interaction}} (M\'{a}laga, Spain) \emph{(\bibinfo{series}{Interacci\'{o}n '21})}. \bibinfo{publisher}{Association for Computing Machinery}, \bibinfo{address}{New York, NY, USA}, Article \bibinfo{articleno}{25}, \bibinfo{numpages}{5}~pages.
\newblock
\showISBNx{9781450375979}
\urldef\tempurl%
\url{https://doi.org/10.1145/3471391.3471421}
\showDOI{\tempurl}


\bibitem[Linxen et~al\mbox{.}(2021b)]%
        {Linxen2021_weird}
\bibfield{author}{\bibinfo{person}{Sebastian Linxen}, \bibinfo{person}{Christian Sturm}, \bibinfo{person}{Florian Br\"{u}hlmann}, \bibinfo{person}{Vincent Cassau}, \bibinfo{person}{Klaus Opwis}, {and} \bibinfo{person}{Katharina Reinecke}.} \bibinfo{year}{2021}\natexlab{b}.
\newblock \showarticletitle{How WEIRD is CHI?}. In \bibinfo{booktitle}{\emph{Proceedings of the 2021 CHI Conference on Human Factors in Computing Systems}} \emph{(\bibinfo{series}{CHI '21})}. \bibinfo{publisher}{Association for Computing Machinery}, \bibinfo{address}{New York, NY, USA}, Article \bibinfo{articleno}{143}, \bibinfo{numpages}{14}~pages.
\newblock
\showISBNx{9781450380966}
\urldef\tempurl%
\url{https://doi.org/10.1145/3411764.3445488}
\showDOI{\tempurl}


\bibitem[Lu et~al\mbox{.}(2021)]%
        {Lu2021}
\bibfield{author}{\bibinfo{person}{Zhicong Lu}, \bibinfo{person}{Chenxinran Shen}, \bibinfo{person}{Jiannan Li}, \bibinfo{person}{Daniel~Shen Hong}, {and} \bibinfo{person}{Wigdor}.} \bibinfo{year}{2021}\natexlab{}.
\newblock \showarticletitle{More Kawaii than a real-person live streamer: Understanding how the otaku community engages with and perceives virtual YouTubers}.
\newblock \bibinfo{journal}{\emph{Proceedings of the 2021 CHI Conference on Human Factors in Computing Systems}}.
\newblock


\bibitem[Lucas et~al\mbox{.}(2018)]%
        {Lucas2018}
\bibfield{author}{\bibinfo{person}{Gale~M. Lucas}, \bibinfo{person}{Jill Boberg}, \bibinfo{person}{David Traum}, \bibinfo{person}{Ron Artstein}, \bibinfo{person}{Jonathan Gratch}, \bibinfo{person}{Alesia Gainer}, \bibinfo{person}{Emmanuel Johnson}, \bibinfo{person}{Anton Leuski}, {and} \bibinfo{person}{Mikio Nakano}.} \bibinfo{year}{2018}\natexlab{}.
\newblock \showarticletitle{Culture, Errors, and Rapport-building Dialogue in Social Agents}.
\newblock \bibinfo{journal}{\emph{Proceedings of the 18th International Conference on Intelligent Virtual Agents}}, \bibinfo{pages}{51--58}.
\newblock
\showISBNx{9781450360135}
\urldef\tempurl%
\url{https://doi.org/10.1145/3267851.3267887}
\showDOI{\tempurl}


\bibitem[Lyubansky and Eidelson(2005)]%
        {Lyubansky2005}
\bibfield{author}{\bibinfo{person}{Mikhail Lyubansky} {and} \bibinfo{person}{Roy~J. Eidelson}.} \bibinfo{year}{2005}\natexlab{}.
\newblock \showarticletitle{Revisiting Du Bois: The Relationship Between African American Double Consciousness and Beliefs About Racial and National Group Experiences}.
\newblock \bibinfo{journal}{\emph{Journal of Black Psychology}} \bibinfo{volume}{31}, \bibinfo{number}{1} (\bibinfo{year}{2005}), \bibinfo{pages}{3--26}.
\newblock
\urldef\tempurl%
\url{https://doi.org/10.1177/0095798404268289}
\showDOI{\tempurl}


\bibitem[MacDorman et~al\mbox{.}(2009)]%
        {MacDorman2009}
\bibfield{author}{\bibinfo{person}{Karl~F MacDorman}, \bibinfo{person}{Sandosh~K Vasudevan}, {and} \bibinfo{person}{Chin-Chang Ho}.} \bibinfo{year}{2009}\natexlab{}.
\newblock \showarticletitle{Does Japan really have robot mania? Comparing attitudes by implicit and explicit measures}.
\newblock \bibinfo{journal}{\emph{AI Soc.}} \bibinfo{volume}{23}, \bibinfo{number}{4} (\bibinfo{date}{July} \bibinfo{year}{2009}), \bibinfo{pages}{485--510}.
\newblock


\bibitem[Mack(2019)]%
        {Mack2019}
\bibfield{author}{\bibinfo{person}{Eric Mack}.} \bibinfo{year}{2019}\natexlab{}.
\newblock \bibinfo{title}{NASA set to hold funeral for silent Mars Opportunity rover}.
\newblock
\newblock
\newblock
\shownote{https://www.cnet.com/science/nasa-looks-set-to-hold-funeral-for-silent-mars-opportunity-rover/}.


\bibitem[Marcus(2013)]%
        {Marcus2013}
\bibfield{author}{\bibinfo{person}{Aaron Marcus}.} \bibinfo{year}{2013}\natexlab{}.
\newblock \showarticletitle{Cross-Cultural User-Experience Design}. In \bibinfo{booktitle}{\emph{SIGGRAPH Asia 2013 Courses}} (Hong Kong) \emph{(\bibinfo{series}{SA '13})}. \bibinfo{publisher}{Association for Computing Machinery}, \bibinfo{address}{New York, NY, USA}, Article \bibinfo{articleno}{8}, \bibinfo{numpages}{31}~pages.
\newblock
\showISBNx{9781450326315}
\urldef\tempurl%
\url{https://doi.org/10.1145/2542266.2542274}
\showDOI{\tempurl}


\bibitem[Marda and Narayan(2021)]%
        {Marda2021}
\bibfield{author}{\bibinfo{person}{Vidushi Marda} {and} \bibinfo{person}{Shivangi Narayan}.} \bibinfo{year}{2021}\natexlab{}.
\newblock \showarticletitle{On the importance of ethnographic methods in AI research}.
\newblock \bibinfo{journal}{\emph{Nature Machine Intelligence}}  \bibinfo{volume}{3} (\bibinfo{year}{2021}), \bibinfo{pages}{187--189}.
\newblock
Issue 3.
\showISSN{2522-5839}
\urldef\tempurl%
\url{https://doi.org/10.1038/s42256-021-00323-0}
\showDOI{\tempurl}


\bibitem[Markus(2016)]%
        {Markus2016}
\bibfield{author}{\bibinfo{person}{Hazel~Rose Markus}.} \bibinfo{year}{2016}\natexlab{}.
\newblock \showarticletitle{What moves people to action? Culture and motivation}.
\newblock \bibinfo{journal}{\emph{Current Opinion in Psychology}}  \bibinfo{volume}{8} (\bibinfo{year}{2016}), \bibinfo{pages}{161--166}.
\newblock
\showISSN{2352-250X}
\urldef\tempurl%
\url{https://doi.org/10.1016/j.copsyc.2015.10.028}
\showDOI{\tempurl}
\newblock
\shownote{Culture}.


\bibitem[Markus and Hamedani(2019)]%
        {Markus2019}
\bibfield{author}{\bibinfo{person}{Hazel~R Markus} {and} \bibinfo{person}{MarYam~G Hamedani}.} \bibinfo{year}{2019}\natexlab{}.
\newblock \showarticletitle{People are culturally shaped shapers: The psychological science of culture and culture change}.
\newblock In \bibinfo{booktitle}{\emph{Handbook of cultural psychology}}, \bibfield{editor}{\bibinfo{person}{D~Cohen} {and} \bibinfo{person}{\&~S Kitayama}} (Eds.). \bibinfo{publisher}{The Guilford Press}, \bibinfo{pages}{11--52}.
\newblock


\bibitem[Markus and Kitayama(1991)]%
        {Markus1991}
\bibfield{author}{\bibinfo{person}{Hazel~R Markus} {and} \bibinfo{person}{Shinobu Kitayama}.} \bibinfo{year}{1991}\natexlab{}.
\newblock \showarticletitle{Culture and the self: Implications for cognition, emotion, and motivation.}
\newblock \bibinfo{journal}{\emph{Psychological Review}}  \bibinfo{volume}{98} (\bibinfo{year}{1991}), \bibinfo{pages}{224--253}.
\newblock
\showISSN{1939-1471(Electronic),0033-295X(Print)}
\urldef\tempurl%
\url{https://doi.org/10.1037/0033-295X.98.2.224}
\showDOI{\tempurl}


\bibitem[Matsuzaki and Lindemann(2016)]%
        {Matsuzaki2016}
\bibfield{author}{\bibinfo{person}{Hironori Matsuzaki} {and} \bibinfo{person}{Gesa Lindemann}.} \bibinfo{year}{2016}\natexlab{}.
\newblock \showarticletitle{The autonomy-safety-paradox of service robotics in Europe and Japan: a comparative analysis}.
\newblock \bibinfo{journal}{\emph{AI \& SOCIETY}}  \bibinfo{volume}{31} (\bibinfo{year}{2016}), \bibinfo{pages}{501--517}.
\newblock
Issue 4.
\showISSN{1435-5655}
\urldef\tempurl%
\url{https://doi.org/10.1007/s00146-015-0630-7}
\showDOI{\tempurl}


\bibitem[Mazlish(1993)]%
        {Mazlish1993}
\bibfield{author}{\bibinfo{person}{ Bruce Mazlish}.} \bibinfo{year}{1993}\natexlab{}.
\newblock \bibinfo{booktitle}{\emph{The Fourth Discontinuity: The Co-evolution of Humans and Machines}}.
\newblock \bibinfo{publisher}{United Kingdom, Yale University Press}.
\newblock


\bibitem[McCarthy and Wright(2005)]%
        {McCarthy2005}
\bibfield{author}{\bibinfo{person}{John McCarthy} {and} \bibinfo{person}{Peter Wright}.} \bibinfo{year}{2005}\natexlab{}.
\newblock \showarticletitle{Putting `felt-life' at the centre of human--computer interaction ({HCI})}.
\newblock \bibinfo{journal}{\emph{Cogn. Technol. Work}} \bibinfo{volume}{7}, \bibinfo{number}{4} (\bibinfo{date}{Nov.} \bibinfo{year}{2005}), \bibinfo{pages}{262--271}.
\newblock


\bibitem[McDonald and Forte(2020)]%
        {McDonald2020}
\bibfield{author}{\bibinfo{person}{Nora McDonald} {and} \bibinfo{person}{Andrea Forte}.} \bibinfo{year}{2020}\natexlab{}.
\newblock \showarticletitle{The Politics of Privacy Theories: Moving from Norms to Vulnerabilities}.
\newblock \bibinfo{journal}{\emph{Proceedings of the 2020 CHI Conference on Human Factors in Computing Systems}}, \bibinfo{pages}{1--14}.
\newblock
\showISBNx{9781450367080}
\urldef\tempurl%
\url{https://doi.org/10.1145/3313831.3376167}
\showDOI{\tempurl}


\bibitem[Miyai(2013)]%
        {Miyai2013}
\bibfield{author}{\bibinfo{person}{Futoshi Miyai}.} \bibinfo{year}{2013}\natexlab{}.
\newblock \bibinfo{title}{North American Audiences' Reaction to Robotics Theater}.
\newblock
\newblock
\newblock
\shownote{Web Magazine Wochi Kochi. From https://www.wochikochi.jp/english/foreign/2013/05/robot-north-american.php}.


\bibitem[Monin and Oppenheimer(2014)]%
        {Monin2014-me}
\bibfield{author}{\bibinfo{person}{Benoît Monin} {and} \bibinfo{person}{Daniel~M. Oppenheimer}.} \bibinfo{year}{2014}\natexlab{}.
\newblock \showarticletitle{The limits of direct replications and the virtues of stimulus sampling}.
\newblock \bibinfo{journal}{\emph{Social Psychology}} \bibinfo{volume}{45}, \bibinfo{number}{4} (\bibinfo{year}{2014}), \bibinfo{pages}{299--300}.
\newblock


\bibitem[Morling et~al\mbox{.}(2002)]%
        {Morling2002}
\bibfield{author}{\bibinfo{person}{Beth Morling}, \bibinfo{person}{Shinobu Kitayama}, {and} \bibinfo{person}{Yuri Miyamoto}.} \bibinfo{year}{2002}\natexlab{}.
\newblock \showarticletitle{Cultural Practices Emphasize Influence in the United States and Adjustment in Japan}.
\newblock \bibinfo{journal}{\emph{Personality and Social Psychology Bulletin}} \bibinfo{volume}{28}, \bibinfo{number}{3} (\bibinfo{year}{2002}), \bibinfo{pages}{311--323}.
\newblock
\urldef\tempurl%
\url{https://doi.org/10.1177/0146167202286003}
\showDOI{\tempurl}
\showeprint{https://doi.org/10.1177/0146167202286003}


\bibitem[Ogbonnaya-Ogburu et~al\mbox{.}(2020)]%
        {Ogbonnaya2020}
\bibfield{author}{\bibinfo{person}{Ihudiya~Finda Ogbonnaya-Ogburu}, \bibinfo{person}{Angela D~R Smith}, \bibinfo{person}{Alexandra To}, {and} \bibinfo{person}{Kentaro Toyama}.} \bibinfo{year}{2020}\natexlab{}.
\newblock \showarticletitle{Critical Race Theory for HCI}.
\newblock \bibinfo{journal}{\emph{Proceedings of the 2020 CHI Conference on Human Factors in Computing Systems}}.
\newblock


\bibitem[O'Leary et~al\mbox{.}(2019)]%
        {OLeary2019}
\bibfield{author}{\bibinfo{person}{Jasper~Tran O'Leary}, \bibinfo{person}{S Zewde}, \bibinfo{person}{J Mankoff}, {and} \bibinfo{person}{D~K Rosner}.} \bibinfo{year}{2019}\natexlab{}.
\newblock \showarticletitle{Who gets to future? Race, representation, and design methods in Africatown}.
\newblock \bibinfo{journal}{\emph{Proceedings of the 2019 CHI Conference on Human Factors in Computing Systems}} (\bibinfo{year}{2019}), \bibinfo{pages}{1--13}.
\newblock


\bibitem[Olson and Olson(2003)]%
        {Olson2003}
\bibfield{author}{\bibinfo{person}{Gary~M Olson} {and} \bibinfo{person}{Judith~S Olson}.} \bibinfo{year}{2003}\natexlab{}.
\newblock \showarticletitle{Human-computer interaction: psychological aspects of the human use of computing}.
\newblock \bibinfo{journal}{\emph{Annu. Rev. Psychol.}}  \bibinfo{volume}{54} (\bibinfo{year}{2003}), \bibinfo{pages}{491--516}.
\newblock
Issue 1.


\bibitem[Oyserman et~al\mbox{.}(2002)]%
        {Oyserman2002-yd}
\bibfield{author}{\bibinfo{person}{Daphna Oyserman}, \bibinfo{person}{Heather~M Coon}, {and} \bibinfo{person}{Markus Kemmelmeier}.} \bibinfo{year}{2002}\natexlab{}.
\newblock \showarticletitle{Rethinking individualism and collectivism: Evaluation of theoretical assumptions and meta-analyses}.
\newblock \bibinfo{journal}{\emph{Psychol. Bull.}} \bibinfo{volume}{128}, \bibinfo{number}{1} (\bibinfo{date}{Jan.} \bibinfo{year}{2002}), \bibinfo{pages}{3--72}.
\newblock


\bibitem[Reinecke and Bernstein(2011)]%
        {Reinecke2011}
\bibfield{author}{\bibinfo{person}{Katharina Reinecke} {and} \bibinfo{person}{Abraham Bernstein}.} \bibinfo{year}{2011}\natexlab{}.
\newblock \showarticletitle{Improving performance, perceived usability, and aesthetics with culturally adaptive user interfaces}.
\newblock \bibinfo{journal}{\emph{ACM Trans. Comput. Hum. Interact.}}  \bibinfo{volume}{18} (\bibinfo{date}{6} \bibinfo{year}{2011}), \bibinfo{pages}{1--29}.
\newblock
Issue 2.


\bibitem[Remagnino and Foresti(2005)]%
        {Remagnino2005}
\bibfield{author}{\bibinfo{person}{Paolo Remagnino} {and} \bibinfo{person}{Gian~Luca Foresti}.} \bibinfo{year}{2005}\natexlab{}.
\newblock \showarticletitle{Ambient Intelligence: A New Multidisciplinary Paradigm}.
\newblock \bibinfo{journal}{\emph{IEEE Transactions on Systems, Man, and Cybernetics - Part A: Systems and Humans}} \bibinfo{volume}{35}, \bibinfo{number}{1} (\bibinfo{year}{2005}), \bibinfo{pages}{1--6}.
\newblock
\urldef\tempurl%
\url{https://doi.org/10.1109/TSMCA.2004.838456}
\showDOI{\tempurl}


\bibitem[Rogers(2012)]%
        {Rogers2012}
\bibfield{author}{\bibinfo{person}{Yvonne Rogers}.} \bibinfo{year}{2012}\natexlab{}.
\newblock \showarticletitle{HCI Theory: Classical, Modern, and Contemporary}.
\newblock \bibinfo{journal}{\emph{Synthesis Lectures on Human-Centered Informatics}}  \bibinfo{volume}{5} (\bibinfo{date}{5} \bibinfo{year}{2012}), \bibinfo{pages}{1--129}.
\newblock
Issue 2.
\showISSN{1946-7680}
\urldef\tempurl%
\url{https://doi.org/10.2200/S00418ED1V01Y201205HCI014}
\showDOI{\tempurl}


\bibitem[Rosseel(2012)]%
        {JSSv048i02}
\bibfield{author}{\bibinfo{person}{Yves Rosseel}.} \bibinfo{year}{2012}\natexlab{}.
\newblock \showarticletitle{lavaan: An R Package for Structural Equation Modeling}.
\newblock \bibinfo{journal}{\emph{Journal of Statistical Software}} \bibinfo{volume}{48}, \bibinfo{number}{2} (\bibinfo{year}{2012}), \bibinfo{pages}{1–36}.
\newblock
\urldef\tempurl%
\url{https://doi.org/10.18637/jss.v048.i02}
\showDOI{\tempurl}


\bibitem[Sakura(2021)]%
        {Sakura2021}
\bibfield{author}{\bibinfo{person}{Osamu Sakura}.} \bibinfo{year}{2021}\natexlab{}.
\newblock \showarticletitle{Examination and international comparison of effectiveness of techno-animism concept in social acceptance of AI and robots}.
\newblock \bibinfo{journal}{\emph{Impact}}  \bibinfo{volume}{2021} (\bibinfo{date}{2} \bibinfo{year}{2021}), \bibinfo{pages}{24--26}.
\newblock
Issue 1.
\showISSN{2398-7073}
\urldef\tempurl%
\url{https://doi.org/10.21820/23987073.2021.1.24}
\showDOI{\tempurl}


\bibitem[Satchell(2008)]%
        {Satchell2008}
\bibfield{author}{\bibinfo{person}{Christine Satchell}.} \bibinfo{year}{2008}\natexlab{}.
\newblock \showarticletitle{Cultural theory and real world design}.
\newblock \bibinfo{journal}{\emph{Proceeding of the twenty-sixth annual CHI conference on Human factors in computing systems - CHI '08}}, \bibinfo{pages}{1593}.
\newblock
\showISBNx{9781605580111}
\urldef\tempurl%
\url{https://doi.org/10.1145/1357054.1357303}
\showDOI{\tempurl}


\bibitem[Savani et~al\mbox{.}(2011)]%
        {Savani2011}
\bibfield{author}{\bibinfo{person}{Krishna Savani}, \bibinfo{person}{Satishchandra Kumar}, \bibinfo{person}{N.~V.~R. Naidu}, {and} \bibinfo{person}{Carol~S. Dweck}.} \bibinfo{year}{2011}\natexlab{}.
\newblock \showarticletitle{Beliefs about emotional residue: The idea that emotions leave a trace in the physical environment.}
\newblock \bibinfo{journal}{\emph{Journal of Personality and Social Psychology}} \bibinfo{volume}{101}, \bibinfo{number}{4} (\bibinfo{year}{2011}), \bibinfo{pages}{684–701}.
\newblock
\showISSN{0022-3514}
\urldef\tempurl%
\url{https://doi.org/10.1037/a0024102}
\showDOI{\tempurl}


\bibitem[Savery et~al\mbox{.}(2021)]%
        {Savery2021}
\bibfield{author}{\bibinfo{person}{Richard Savery}, \bibinfo{person}{Lisa Zahray}, {and} \bibinfo{person}{Gil Weinberg}.} \bibinfo{year}{2021}\natexlab{}.
\newblock \bibinfo{booktitle}{\emph{Shimon Sings-Robotic Musicianship Finds Its Voice}}.
\newblock \bibinfo{publisher}{Springer International Publishing}, \bibinfo{address}{Cham}, \bibinfo{pages}{823--847}.
\newblock
\showISBNx{978-3-030-72116-9}
\urldef\tempurl%
\url{https://doi.org/10.1007/978-3-030-72116-9_29}
\showDOI{\tempurl}


\bibitem[Schwarz(1999)]%
        {Schwarz1999}
\bibfield{author}{\bibinfo{person}{Norbert Schwarz}.} \bibinfo{year}{1999}\natexlab{}.
\newblock \showarticletitle{Self-reports: How the questions shape the answers}.
\newblock \bibinfo{journal}{\emph{Am. Psychol.}} \bibinfo{volume}{54}, \bibinfo{number}{2} (\bibinfo{date}{Feb.} \bibinfo{year}{1999}), \bibinfo{pages}{93--105}.
\newblock


\bibitem[Scissors et~al\mbox{.}(2011)]%
        {Scissors2011}
\bibfield{author}{\bibinfo{person}{Lauren Scissors}, \bibinfo{person}{N.~Sadat Shami}, \bibinfo{person}{Tatsuya Ishihara}, \bibinfo{person}{Steven Rohall}, {and} \bibinfo{person}{Shin Saito}.} \bibinfo{year}{2011}\natexlab{}.
\newblock \showarticletitle{Real-time collaborative editing behavior in USA and Japanese distributed teams}.
\newblock \bibinfo{journal}{\emph{Proceedings of the SIGCHI Conference on Human Factors in Computing Systems}}, \bibinfo{pages}{1119--1128}.
\newblock
\showISBNx{9781450302289}
\urldef\tempurl%
\url{https://doi.org/10.1145/1978942.1979109}
\showDOI{\tempurl}


\bibitem[Seymour and Kleek(2020)]%
        {Seymour2020}
\bibfield{author}{\bibinfo{person}{William Seymour} {and} \bibinfo{person}{Max~Van Kleek}.} \bibinfo{year}{2020}\natexlab{}.
\newblock \showarticletitle{Does Siri have a soul? Exploring voice assistants through Shinto design fictions}.
\newblock \bibinfo{journal}{\emph{Extended Abstracts of the 2020 CHI Conference on Human Factors in Computing Systems}}.
\newblock
\showISBNx{9781450368193}
\urldef\tempurl%
\url{https://doi.org/10.1145/3334480.3381809}
\showDOI{\tempurl}


\bibitem[Shneiderman et~al\mbox{.}(2002)]%
        {Shneiderman2002}
\bibfield{author}{\bibinfo{person}{Ben Shneiderman}, \bibinfo{person}{Stuart Card}, \bibinfo{person}{Donald~A. Norman}, \bibinfo{person}{Marilyn Tremaine}, {and} \bibinfo{person}{M.~Mitchell Waldrop}.} \bibinfo{year}{2002}\natexlab{}.
\newblock \showarticletitle{CHI@20: fighting our way from marginality to power}.
\newblock \bibinfo{journal}{\emph{CHI '02 extended abstracts on Human factors in computing systems - CHI '02}}, \bibinfo{pages}{688--691}.
\newblock
\showISBNx{1581134541}
\urldef\tempurl%
\url{https://doi.org/10.1145/506443.506548}
\showDOI{\tempurl}


\bibitem[Soro et~al\mbox{.}(2019a)]%
        {Soro2019}
\bibfield{author}{\bibinfo{person}{Alessandro Soro}, \bibinfo{person}{Margot Brereton}, \bibinfo{person}{Laurianne Sitbon}, \bibinfo{person}{Aloha~Hufana Ambe}, \bibinfo{person}{Jennyfer~Lawrence Taylor}, {and} \bibinfo{person}{Cara Wilson}.} \bibinfo{year}{2019}\natexlab{a}.
\newblock \showarticletitle{Beyond Independence}.
\newblock \bibinfo{journal}{\emph{Proceedings of the 31st Australian Conference on Human-Computer-Interaction}}, \bibinfo{pages}{149--160}.
\newblock
\showISBNx{9781450376969}
\urldef\tempurl%
\url{https://doi.org/10.1145/3369457.3369470}
\showDOI{\tempurl}


\bibitem[Soro et~al\mbox{.}(2019b)]%
        {Soro2019_designing_past}
\bibfield{author}{\bibinfo{person}{Alessandro Soro}, \bibinfo{person}{Jennyfer~Lawrence Taylor}, {and} \bibinfo{person}{Margot Brereton}.} \bibinfo{year}{2019}\natexlab{b}.
\newblock \showarticletitle{Designing the past}.
\newblock \bibinfo{journal}{\emph{Extended Abstracts of the 2019 CHI Conference on Human Factors in Computing Systems}}.
\newblock


\bibitem[Spatola et~al\mbox{.}(2022)]%
        {Spatola2022}
\bibfield{author}{\bibinfo{person}{Nicolas Spatola}, \bibinfo{person}{Serena Marchesi}, {and} \bibinfo{person}{Agnieszka Wykowska}.} \bibinfo{year}{2022}\natexlab{}.
\newblock \showarticletitle{Different models of anthropomorphism across cultures and ontological limits in current frameworks the integrative framework of anthropomorphism}.
\newblock \bibinfo{journal}{\emph{Frontiers in Robotics and AI}}  \bibinfo{volume}{9} (\bibinfo{date}{8} \bibinfo{year}{2022}).
\newblock
\showISSN{2296-9144}
\urldef\tempurl%
\url{https://doi.org/10.3389/frobt.2022.863319}
\showDOI{\tempurl}


\bibitem[Steinfeld et~al\mbox{.}(2009)]%
        {Steinfeld2009}
\bibfield{author}{\bibinfo{person}{Aaron Steinfeld}, \bibinfo{person}{Odest~Chadwicke Jenkins}, {and} \bibinfo{person}{Brian Scassellati}.} \bibinfo{year}{2009}\natexlab{}.
\newblock \showarticletitle{The Oz of Wizard: Simulating the Human for Interaction Research}. In \bibinfo{booktitle}{\emph{Proceedings of the 4th ACM/IEEE International Conference on Human Robot Interaction}} (La Jolla, California, USA) \emph{(\bibinfo{series}{HRI '09})}. \bibinfo{publisher}{Association for Computing Machinery}, \bibinfo{address}{New York, NY, USA}, \bibinfo{pages}{101–108}.
\newblock
\showISBNx{9781605584041}
\urldef\tempurl%
\url{https://doi.org/10.1145/1514095.1514115}
\showDOI{\tempurl}


\bibitem[Suchman(1987)]%
        {suchman1987}
\bibfield{author}{\bibinfo{person}{Lucy~A. Suchman}.} \bibinfo{year}{1987}\natexlab{}.
\newblock \bibinfo{booktitle}{\emph{Plans and situated actions: The problem of human-machine communication}}.
\newblock \bibinfo{publisher}{Cambridge university press}.
\newblock


\bibitem[Talhelm et~al\mbox{.}(2014)]%
        {Talhelm2014}
\bibfield{author}{\bibinfo{person}{T. Talhelm}, \bibinfo{person}{X. Zhang}, \bibinfo{person}{S. Oishi}, \bibinfo{person}{C. Shimin}, \bibinfo{person}{D. Duan}, \bibinfo{person}{X. Lan}, {and} \bibinfo{person}{S. Kitayama}.} \bibinfo{year}{2014}\natexlab{}.
\newblock \showarticletitle{Large-Scale Psychological Differences Within China Explained by Rice Versus Wheat Agriculture}.
\newblock \bibinfo{journal}{\emph{Science}} \bibinfo{volume}{344}, \bibinfo{number}{6184} (\bibinfo{year}{2014}), \bibinfo{pages}{603--608}.
\newblock
\urldef\tempurl%
\url{https://doi.org/10.1126/science.1246850}
\showDOI{\tempurl}
\showeprint{https://www.science.org/doi/pdf/10.1126/science.1246850}


\bibitem[Taylor(2011)]%
        {Taylor2011}
\bibfield{author}{\bibinfo{person}{Alex~S. Taylor}.} \bibinfo{year}{2011}\natexlab{}.
\newblock \showarticletitle{Out there}.
\newblock \bibinfo{journal}{\emph{Proceedings of the SIGCHI Conference on Human Factors in Computing Systems}}, \bibinfo{pages}{685--694}.
\newblock
\showISBNx{9781450302289}
\urldef\tempurl%
\url{https://doi.org/10.1145/1978942.1979042}
\showDOI{\tempurl}


\bibitem[Triandis and Gelfand(1998)]%
        {Triandis1998}
\bibfield{author}{\bibinfo{person}{Harry~C Triandis} {and} \bibinfo{person}{Michele~J Gelfand}.} \bibinfo{year}{1998}\natexlab{}.
\newblock \showarticletitle{Converging measurement of horizontal and vertical individualism and collectivism}.
\newblock \bibinfo{journal}{\emph{J. Pers. Soc. Psychol.}} \bibinfo{volume}{74}, \bibinfo{number}{1} (\bibinfo{date}{Jan.} \bibinfo{year}{1998}), \bibinfo{pages}{118--128}.
\newblock


\bibitem[Tsai et~al\mbox{.}(2007)]%
        {tsai2007influence}
\bibfield{author}{\bibinfo{person}{Jeanne~L. Tsai}, \bibinfo{person}{Felicity~F. Miao}, \bibinfo{person}{Emma Seppala}, \bibinfo{person}{Helene~H. Fung}, {and} \bibinfo{person}{Dannii~Y. Yeung}.} \bibinfo{year}{2007}\natexlab{}.
\newblock \showarticletitle{Influence and adjustment goals: Sources of cultural differences in ideal affect}.
\newblock \bibinfo{journal}{\emph{Journal of Personality and Social Psychology}} \bibinfo{volume}{92}, \bibinfo{number}{6} (\bibinfo{year}{2007}), \bibinfo{pages}{1102--1117}.
\newblock
\urldef\tempurl%
\url{https://doi.org/10.1037/0022-3514.92.6.1102}
\showDOI{\tempurl}


\bibitem[Uriu and Okude(2010)]%
        {Uriu2010}
\bibfield{author}{\bibinfo{person}{Daisuke Uriu} {and} \bibinfo{person}{Naohito Okude}.} \bibinfo{year}{2010}\natexlab{}.
\newblock \showarticletitle{ThanatoFenestra}.
\newblock \bibinfo{journal}{\emph{Proceedings of the 8th ACM Conference on Designing Interactive Systems - DIS '10}}, \bibinfo{pages}{422}.
\newblock
\showISBNx{9781450301039}
\urldef\tempurl%
\url{https://doi.org/10.1145/1858171.1858253}
\showDOI{\tempurl}


\bibitem[Vatrapu(2010)]%
        {Vatrapu2010}
\bibfield{author}{\bibinfo{person}{Ravi~K. Vatrapu}.} \bibinfo{year}{2010}\natexlab{}.
\newblock \showarticletitle{Explaining culture}.
\newblock \bibinfo{journal}{\emph{Proceedings of the 3rd international conference on Intercultural collaboration - ICIC '10}}, \bibinfo{pages}{111}.
\newblock
\showISBNx{9781450301084}
\urldef\tempurl%
\url{https://doi.org/10.1145/1841853.1841871}
\showDOI{\tempurl}


\bibitem[Walker(2018)]%
        {Walker2018}
\bibfield{author}{\bibinfo{person}{Sheena~Myong Walker}.} \bibinfo{year}{2018}\natexlab{}.
\newblock \showarticletitle{Empirical Study of the Application of Double-Consciousness Among African-American Men}.
\newblock \bibinfo{journal}{\emph{Journal of African American Studies}} \bibinfo{volume}{22}, \bibinfo{number}{2–3} (\bibinfo{date}{Aug.} \bibinfo{year}{2018}), \bibinfo{pages}{205–217}.
\newblock
\showISSN{1936-4741}
\urldef\tempurl%
\url{https://doi.org/10.1007/s12111-018-9404-x}
\showDOI{\tempurl}


\bibitem[Wang et~al\mbox{.}(2010)]%
        {Wang2010}
\bibfield{author}{\bibinfo{person}{Lin Wang}, \bibinfo{person}{Pei-Luen~Patrick Rau}, \bibinfo{person}{Vanessa Evers}, \bibinfo{person}{Benjamin~Krisper Robinson}, {and} \bibinfo{person}{Pamela Hinds}.} \bibinfo{year}{2010}\natexlab{}.
\newblock \showarticletitle{When in Rome: The role of culture \& context in adherence to robot recommendations}.
\newblock \bibinfo{journal}{\emph{2010 5th ACM/IEEE International Conference on Human-Robot Interaction (HRI)}}.
\newblock


\bibitem[Wang(2004)]%
        {Wang2004-fn}
\bibfield{author}{\bibinfo{person}{Qi Wang}.} \bibinfo{year}{2004}\natexlab{}.
\newblock \showarticletitle{The emergence of cultural self-constructs: autobiographical memory and self-description in European American and Chinese children}.
\newblock \bibinfo{journal}{\emph{Dev. Psychol.}} \bibinfo{volume}{40}, \bibinfo{number}{1} (\bibinfo{date}{Jan.} \bibinfo{year}{2004}), \bibinfo{pages}{3--15}.
\newblock


\bibitem[Warne(2014)]%
        {warne2014primer}
\bibfield{author}{\bibinfo{person}{Russell~T. Warne}.} \bibinfo{year}{2014}\natexlab{}.
\newblock \showarticletitle{A Primer on Multivariate Analysis of Variance (MANOVA) for Behavioral Scientists}.
\newblock \bibinfo{journal}{\emph{Practical Assessment, Research \& Evaluation}}  \bibinfo{volume}{19} (\bibinfo{year}{2014}).
\newblock


\bibitem[Weng et~al\mbox{.}(2019)]%
        {Weng2019}
\bibfield{author}{\bibinfo{person}{Yueh-Hsuan Weng}, \bibinfo{person}{Yasuhisa Hirata}, \bibinfo{person}{Osamu Sakura}, {and} \bibinfo{person}{Yusuke Sugahara}.} \bibinfo{year}{2019}\natexlab{}.
\newblock \showarticletitle{The Religious Impacts of Taoism on Ethically Aligned Design in HRI}.
\newblock \bibinfo{journal}{\emph{International Journal of Social Robotics}}  \bibinfo{volume}{11} (\bibinfo{year}{2019}), \bibinfo{pages}{829--839}.
\newblock
Issue 5.
\showISSN{1875-4805}
\urldef\tempurl%
\url{https://doi.org/10.1007/s12369-019-00594-z}
\showDOI{\tempurl}


\bibitem[White and Katsuno(2021)]%
        {White2021}
\bibfield{author}{\bibinfo{person}{Daniel White} {and} \bibinfo{person}{Hirofumi Katsuno}.} \bibinfo{year}{2021}\natexlab{}.
\newblock \showarticletitle{Toward an Affective Sense of Life: Artificial Intelligence, Animacy, and Amusement at a Robot Pet Memorial Service in Japan}.
\newblock \bibinfo{journal}{\emph{Cultural Anthropology}}  \bibinfo{volume}{36} (\bibinfo{date}{5} \bibinfo{year}{2021}).
\newblock
Issue 2.
\showISSN{1548-1360}
\urldef\tempurl%
\url{https://doi.org/10.14506/ca36.2.03}
\showDOI{\tempurl}


\bibitem[Woodruff et~al\mbox{.}(2018)]%
        {Woodruff2018}
\bibfield{author}{\bibinfo{person}{Allison Woodruff}, \bibinfo{person}{Sarah~E. Fox}, \bibinfo{person}{Steven Rousso-Schindler}, {and} \bibinfo{person}{Jeffrey Warshaw}.} \bibinfo{year}{2018}\natexlab{}.
\newblock \showarticletitle{A Qualitative Exploration of Perceptions of Algorithmic Fairness}. In \bibinfo{booktitle}{\emph{Proceedings of the 2018 CHI Conference on Human Factors in Computing Systems}} (Montreal QC, Canada) \emph{(\bibinfo{series}{CHI '18})}. \bibinfo{publisher}{Association for Computing Machinery}, \bibinfo{address}{New York, NY, USA}, \bibinfo{pages}{1–14}.
\newblock
\showISBNx{9781450356206}
\urldef\tempurl%
\url{https://doi.org/10.1145/3173574.3174230}
\showDOI{\tempurl}


\bibitem[Yang et~al\mbox{.}(2011)]%
        {Yang2011}
\bibfield{author}{\bibinfo{person}{Jiang Yang}, \bibinfo{person}{Mark~S. Ackerman}, {and} \bibinfo{person}{Lada~A. Adamic}.} \bibinfo{year}{2011}\natexlab{}.
\newblock \showarticletitle{Virtual gifts and guanxi}.
\newblock \bibinfo{journal}{\emph{Proceedings of the ACM 2011 conference on Computer supported cooperative work - CSCW '11}}, \bibinfo{pages}{45}.
\newblock
\showISBNx{9781450305563}
\urldef\tempurl%
\url{https://doi.org/10.1145/1958824.1958832}
\showDOI{\tempurl}


\bibitem[Yang et~al\mbox{.}(2013)]%
        {Yang2013}
\bibfield{author}{\bibinfo{person}{Jiang Yang}, \bibinfo{person}{Scott Counts}, \bibinfo{person}{Meredith~Ringel Morris}, {and} \bibinfo{person}{Aaron Hoff}.} \bibinfo{year}{2013}\natexlab{}.
\newblock \showarticletitle{Microblog credibility perceptions}.
\newblock \bibinfo{journal}{\emph{Proceedings of the 2013 conference on Computer supported cooperative work - CSCW '13}}, \bibinfo{pages}{575}.
\newblock
\showISBNx{9781450313315}
\urldef\tempurl%
\url{https://doi.org/10.1145/2441776.2441841}
\showDOI{\tempurl}


\bibitem[Youichiro(2020)]%
        {Youichiro2020}
\bibfield{author}{\bibinfo{person}{Miyake Youichiro}.} \bibinfo{year}{2020}\natexlab{}.
\newblock \bibinfo{booktitle}{\emph{When Artificial Intelligence becomes life}}.
\newblock \bibinfo{publisher}{PLANETS}.
\newblock


\bibitem[Zeeberg(2020)]%
        {Zeeberg2020}
\bibfield{author}{\bibinfo{person}{Amos Zeeberg}.} \bibinfo{year}{2020}\natexlab{}.
\newblock \bibinfo{title}{What we can learn about robots from Japan}.
\newblock \bibinfo{howpublished}{\url{https://www.bbc.com/future/article/20191220-what-we-can-learn-about-robots-from-japan}}.
\newblock
\newblock
\shownote{Accessed: 2023-9-12}.


\bibitem[Zhao and Jiang(2011)]%
        {Zhao2011}
\bibfield{author}{\bibinfo{person}{Chen Zhao} {and} \bibinfo{person}{Gonglue Jiang}.} \bibinfo{year}{2011}\natexlab{}.
\newblock \showarticletitle{Cultural differences on visual self-presentation through social networking site profile images}.
\newblock \bibinfo{journal}{\emph{Proceedings of the SIGCHI Conference on Human Factors in Computing Systems}}, \bibinfo{pages}{1129--1132}.
\newblock
\showISBNx{9781450302289}
\urldef\tempurl%
\url{https://doi.org/10.1145/1978942.1979110}
\showDOI{\tempurl}


\end{thebibliography}










\end{document}